\documentclass[10pt,letterpaper,twocolumn,english,journal]{IEEEtran}
\usepackage[T1]{fontenc}
\usepackage{float}
\usepackage{amsmath}
\usepackage{amsthm}
\usepackage{graphicx}

\makeatletter

\floatstyle{ruled}
\newfloat{algorithm}{tbp}{loa}
\providecommand{\algorithmname}{Algorithm}
\floatname{algorithm}{\protect\algorithmname}

\theoremstyle{plain}
\newtheorem{thm}{\protect\theoremname}
\theoremstyle{plain}
\newtheorem{lem}[thm]{\protect\lemmaname}

\ifCLASSOPTIONcompsoc
\usepackage[caption=false,font=normalsize,labelfont=sf,textfont=sf]{subfig}
\else
\usepackage[caption=false,font=footnotesize]{subfig}
\fi

\usepackage{cite}
\usepackage{url}
\usepackage{algorithmic}
\usepackage{algorithm}
\usepackage{caption}
\newtheorem{Thm}{Theorem}
\newtheorem{Prop}{Proposition}

\makeatother

\usepackage{babel}
\providecommand{\lemmaname}{Lemma}
\providecommand{\theoremname}{Theorem}

\begin{document}
\title{When NOMA Meets AIGC: Enhanced Wireless Federated Learning}
\author{Ding Xu,~\IEEEmembership{Senior Member,~IEEE}, Lingjie Duan,~\IEEEmembership{Senior Member,~IEEE},
and Hongbo Zhu,~\IEEEmembership{Member,~IEEE}\thanks{Ding Xu is with the Jiangsu Key Laboratory of Wireless Communications,
Nanjing University of Posts and Telecommunications, Nanjing 210003,
China (E-mail: xuding@ieee.org). He is also with the Pillar of Engineering
Systems and Design, Singapore University of Technology and Design,
Singapore 487372, Singapore.}\thanks{Lingjie Duan is with the Pillar of Engineering Systems and Design,
Singapore University of Technology and Design, Singapore 487372, Singapore
(E-mail: lingjie\_duan@sutd.edu.sg).}\thanks{Hongbo Zhu is with the Jiangsu Key Laboratory of Wireless Communications,
Nanjing University of Posts and Telecommunications, Nanjing 210003,
China (E-mail: zhuhb@njupt.edu.cn).}}
\maketitle
\begin{abstract}
Wireless federated learning (WFL) enables devices to collaboratively
train a global model via local model training, uploading and aggregating.
However, WFL faces the data scarcity/heterogeneity problem (i.e.,
data are limited and unevenly distributed among devices) that degrades
the learning performance. In this regard, artificial intelligence
generated content (AIGC) can synthesize various types of data to compensate
for the insufficient local data. Nevertheless, downloading synthetic
data or uploading local models iteratively takes a lot of time, especially
for a large amount of devices. To address this issue, we propose to
leverage non-orthogonal multiple access (NOMA) to achieve efficient
synthetic data and local model transmission. This paper is the first
to combine AIGC and NOMA with WFL to maximally enhance the learning
performance. For the proposed NOMA+AIGC-enhanced WFL, the problem
of jointly optimizing the synthetic data distribution, two-way communication
and computation resource allocation to minimize the global learning
error is investigated. The problem belongs to NP-hard mixed integer
nonlinear programming, whose optimal solution is intractable to find.
We first employ the block coordinate descent method to decouple the
complicated-coupled variables, and then resort to our analytical method
to derive an efficient low-complexity local optimal solution with
partial closed-form results. Extensive simulations validate the superiority
of the proposed scheme compared to the existing and benchmark schemes
such as the frequency/time division multiple access based AIGC-enhanced
schemes.
\end{abstract}

\begin{IEEEkeywords}
Non-orthogonal multiple access, artificial intelligence generated
content, wireless federated learning, synthetic data.
\end{IEEEkeywords}

\section{Introduction}

\bstctlcite{IEEEexample:BSTcontrol}

Nowadays, due to the explosive growth of Internet of Things (IoT)
devices, the data collected by IoT devices are needed to be analyzed
using machine learning techniques to support various IoT applications
such as augmented reality and virtual reality \cite{javed2018internet}.
However, because of the limited communication resources, it is unbearable
to transmit all of the collected data to a data center to perform
centralized machine learning. Luckily, the computing capability of
devices is surging due to the fast development of chip technology,
and it motivates to implement distributed machine learning that lets
each device train a learning model locally using its collected data.
Wireless federated learning (WFL) is one of the most famous distributed
learning framework that allows devices to collaboratively learn a
global model while protecting the data privacy \cite{khan2021federated}.
In WFL, devices train local models using local data independently
based on a received global model, then the trained local models are
transmitted to the WFL server for aggregating to a global model, while
the new global model is redistributed back to the devices, and the
process repeated until the global model converges. Therefore, wireless
transmission schemes affect the WFL performance greatly and need to
be tailored for WFL.

Meanwhile, due to data scarcity and heterogeneity, data are limited
and unevenly distributed among devices and some important portions
of data are missing locally at some devices, which leads to poor learning
convergence accuracy \cite{wang2020machine}. Therefore, to improve
global model accuracy, some measures have to be taken. For example,
device selection can be performed to select proper devices for global
model aggregation \cite{khan2021federated}. However, this will introduce
the fairness issue. Another method is to let devices collect the missing
portions of data. However, due to physical limitations, sometimes
it is difficult for the devices to collect the missing portions of
local data by themselves. Even if collecting data is possible, it
will certainly introduce much more latency and energy consumption,
which is not desirable for energy-limited devices and delay-sensitive
services. In this regard, artificial intelligence generated content
(AIGC) \cite{celik2024dawn}, which is a promising technology for
synthesizing data, can be adopted to generate synthetic data for device
local training. Specifically, AIGC can automatically create various
data such as texts, images, and videos as the devices request \cite{karapantelakis2024generative,xu2024unleashing},
and can save time and resources that may otherwise be spent on the
data collection. Since AIGC usually requires high computation capability,
it can be deployed at an AIGC server in the cloud computing center,
and devices can request the server for synthesizing specific data
and then download the synthetic data for local model training.

\subsection{Related Works}

In the context of WFL networks, many works such as \cite{chen2020joint,yang2021energy,bouzinis2023wireless,zhang2023joint,zhao2024economic}
have studied various important problems regarding the implementation
of WFL. Specifically, in \cite{chen2020joint}, an algorithm for jointly
optimizing the learning, radio resource allocation and device selection
based on the Hungarian method was proposed to minimize the WFL loss
function. In \cite{yang2021energy}, the problem of joint learning
and communication resource optimization to minimize the total energy
consumption was investigated, and an iterative algorithm was proposed
to achieve a local optimal solution. In \cite{bouzinis2023wireless},
the computation and the communication resources as well as the number
of the local model parameter quantization bits were jointly optimized
to minimize the WFL convergence time. In \cite{zhang2023joint}, an
iterative algorithm for jointly scheduling devices, local iterations
and radio resources was developed based on the pointer network embedded
deep reinforcement learning method and the breadth-first search method.
In \cite{zhao2024economic}, an incentive mechanism based on the Stackelberg
game was designed to motivate the devices to participate in collaborative
model training.

In the above works, \cite{chen2020joint,yang2021energy,zhang2023joint,zhao2024economic}
adopted the frequency division multiple access (FDMA) wireless transmission
scheme, and \cite{bouzinis2023wireless} adopted the time division
multiple access (TDMA) wireless transmission scheme. Both FDMA and
TDMA belong to orthogonal multiple access (OMA) transmission scheme.
However, OMA suffers from low transmission efficiency \cite{xu2021sum,liu2022evolution}.
To fully realize the potential capability of WFL, more efficient transmission
schemes other than OMA are needed. In this context, non-orthogonal
multiple access (NOMA) \cite{liu2022evolution,xu2024device}, which
supports multiple concurrent transmissions on the same spectrum band,
is a promising technique to realize high transmission efficiency in
WFL networks. Particularly, devices can upload their local models
simultaneously using the superposition coding, and the WFL server
can decode different device signals by using the successive interference
cancellation (SIC) technique. 

There have been some works on NOMA-enhanced WFL networks. Specifically,
in \cite{wu2022non}, devices were assumed to be wirelessly powered
by the base station (BS), and the problem of joint optimization of
the NOMA communication and computation resources to minimize the system-wise
cost was investigated. A layered algorithm based on the monotonic
optimization was developed to solve the joint optimization problem.
In \cite{al2022energy}, the problem of jointly optimizing device
scheduling, transmit power and computation frequency allocation in
a relay-assisted NOMA-enhanced WFL network to minimize the energy
consumption was solved by graph theory. In \cite{li2023multi}, devices
were grouped into different NOMA groups for local model uploading,
and the transmit power and bandwidth of NOMA groups were jointly optimized
to maximize the system convergence metric based on convex optimization.
 In \cite{wu2024joint}, the joint optimization of device selection
and resource allocation to minimize the total training latency in
a NOMA-enhanced WFL network was carried by the monotonicity analysis
and dual decomposition method.

Meanwhile, AIGC in wireless networks has attracted a lot of attention
recently, such as the AIGC-enhanced semantic communications \cite{liu2024semantic,cheng2023wireless,wu2024cddm},
blockchain-enabled AIGC \cite{liu2024blockchain}, distributed AIGC
\cite{du2023exploring}, and AIGC service provider selection \cite{du2024diffusion}.
AIGC has also been integrated with WFL in recent works \cite{huang2024federated,li2024filling}.
Particularly, the work in \cite{huang2024federated} applied WFL to
achieve efficient AIGC, and presented WFL-based techniques for AIGC
to generate diverse and personalized contents. While \cite{huang2024federated}
focused on how WFL can empower the AIGC, the work in \cite{li2024filling}
adopted AIGC to empower WFL. Specifically, in \cite{li2024filling},
AIGC was proposed to generate more training data for devices to minimize
the device energy consumption under the learning performance constraint.

\subsection{Motivation and Contributions}

Since devices are heterogeneous and the data are limited and unevenly
distributed, AIGC can be used to generate the specified data missing
at the devices for more efficient local training and improved global
convergence performance in WFL. In this regard, the pioneer work \cite{li2024filling}
has considered to use AIGC to enhance the WFL performance. However,
the work \cite{li2024filling} ignored the procedure of synthetic
data downloading from the AIGC server to the devices, the time of
which cannot be neglected in practice. In addition, the work \cite{li2024filling}
used a low-efficient FDMA transmission scheme for local model uploading,
which is inadequate for a large amount of devices. The above research
gaps motivate the work in this paper to use the highly efficient NOMA
transmission scheme both for synthetic data downloading and local
model uploading in AIGC-enhanced WFL networks, and jointly design
the synthetic data distribution, two-way communication resource and
computation resource allocation to maximize the learning performance.
Note that jointly designing resource optimization policy in NOMA+AIGC-enhanced
WFL networks is non-trivial, since the optimization variables are
highly coupled in such complicated non-convex optimization problems,
while the SIC decoding policy in NOMA can further complicate the optimization.

The main contributions of this paper are summarized as follows:

(1) We are the first to propose a NOMA+AIGC-enhanced WFL system model,
where devices can download synthetic data from the AIGC server based
on NOMA, then train the local models based on the local data and synthetic
data, and upload the local models to the WFL server based on NOMA
for global model aggregation. The problem of jointly optimizing the
synthetic data allocation, the time allocation, the transmit power
allocation of the BS and the devices, the SIC decoding order, and
the computing frequency allocation, is formulated with the objective
of minimizing the global learning error, under various system constraints.

(2) We propose an efficient low-complexity algorithm with partial
closed-form results to achieve a local optimal solution to the problem.
First, we analytically derive the closed-form optimal computing frequency
allocation to simplify the problem. Then, the block coordinate descent
(BCD) method is adopted to decouple the complex problem into two simpler
subproblems. The first subproblem optimizes the synthetic data allocation
and the time allocation, where the optimal time allocation is analytical
obtained in closed form, and the optimal synthetic data allocation
is obtained via convex optimization. The second subproblem optimizes
the transmit power allocation of the BS, the transmit power of the
devices, and the SIC decoding order, where the optimal transmit power
allocation of the BS and the transmit power of the devices are analytically
obtained in a recursive form, and the optimal SIC decoding order is
analytically obtained in closed form.

(3) Extensive simulation results are illustrated to demonstrate the
superiority of the proposed NOMA+AIGC-enhanced scheme. The results
show the advantages of combining NOMA and AIGC with WFL. It is shown
that the proposed scheme outperforms the existing and benchmark schemes
such as the FDMA or TDMA+AIGC-enhanced schemes, in terms of global
learning accuracy, under various system configurations. Particularly,
the learning performance improvement of our proposed scheme is more
obvious when the data synthesizing capability of the AIGC server is
stronger, the maximum latency requirement is more stringent, and the
number of devices is larger.

The remainder of this paper is organized as follows. Section \ref{sec:System-Model}
presents the system model of the proposed NOMA+AIGC-enhanced WFL network.
Section \ref{sec:Proposed-Algorithm-forall} formulates the optimization
problem of maximizing the learning performance and presents an algorithm
framework to solve the optimization problem. Section \ref{sec:Proposed-Algorithm-D}
and \ref{sec:Proposed-Algorithm-p} presents algorithms to optimally
solve the two subproblems. Section \ref{sec:Simulation-Results} provides
simulation examples to demonstrate the effectiveness of the proposed
NOMA+AIGC-enhanced scheme. Finally, the paper is concluded in Section
\ref{sec:Conclusions}.

\section{System Model\label{sec:System-Model}}

\begin{figure}[!t]
\centering\includegraphics[width=0.75\columnwidth]{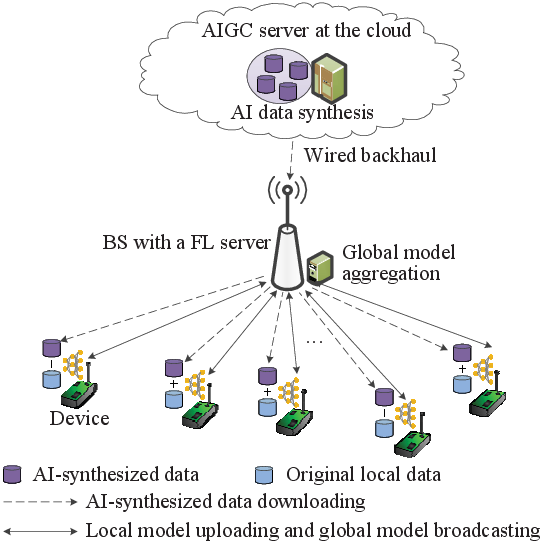}

\caption{NOMA+AIGC-enhanced WFL system model.\label{fig:System-model}}
\end{figure}

We consider a NOMA+AIGC-enhanced WFL network consisting of an AIGC
server in the cloud, a BS and $K$ devices, as shown in Fig. \ref{fig:System-model}.
The BS is equipped with a WFL server for simple computation of global
model aggregation, and connects with the upper AIGC server via high-speed
wired backhaul. The AIGC server is assumed to be deployed at the cloud
computing center and equipped with powerful computation capacity for
data synthesis. Each device has a set of original local dataset for
local model training. Let $D_{k}^{\mathrm{loc}}$ denote the number
of samples in the local dataset of device $k.$ In order to improve
the learning model accuracy, AIGC is adopted to synthesize the data
required by the devices for performing local training. Specifically,
each device can send the data synthesis request to the AIGC server
indicating the specific data distributions/properties. After the AIGC
server receives the requests from the devices, it will synthesize
the requested data and then send the synthetic data to the BS and
then the devices for local model training. Let $D_{k}^{\mathrm{gen}}$
denote the amount of synthetic data samples that are generated by
a pre-trained AI generative model and are downloaded from the AIGC
server to the device $k$. After the AIGC server pushes the synthetic
data to the devices, the devices can train the local models based
on both the local dataset and the synthetic dataset. 

\subsection{AIGC-Enhanced WFL Model}

The whole AIGC-enhanced WFL procedure consists of three phases \cite{li2024filling}.
In the first phase, all the devices send the data synthesis request
to the AIGC server indicating the specific data distributions/properties,
and then the AIGC server synthesizes the data required by all the
devices. Due to data scarcity/heterogeneity, the local data available
at the devices may lack particular types of data, which may hinder
the convergence of the global model. Thus, the synthetic data can
compensate for the missing portions of local data to improve the learning
convergence \cite{li2024filling}. We assume that due to physical
limitations, the AIGC server can generate a maximum amount of $D^{\mathrm{gen}}$
synthetic data \cite{xu2024unleashing}. Since the synthetic data
received by all the devices cannot exceed the maximum amount to bear
by the AIGC server, we have the following synthetic data constraint
as given by 
\begin{equation}
\sum_{k=1}^{K}D_{k}^{\mathrm{gen}}\leq D^{\mathrm{gen}}.
\end{equation}
In the second phase, the synthetic data are transmitted from the AIGC
server in the cloud to the BS and then the devices with time $T^{\mathrm{down}}$.
Since the AIGC server connects with the BS via high-speed wired backhaul,
the time for transmission the synthetic data from the AIGC server
to the BS is ignored. In the third phase, the WFL server first broadcasts
the initial global model to the devices, then each device trains the
local model using the local dataset and the synthetic dataset, and
finally uploads the trained local model to the WFL server for global
model aggregation and update. We assume that the number of global
model iterations is fixed at $N.$ Let $D^{\mathrm{mod}},$ $T^{\mathrm{syn}},$
$T^{\mathrm{br}},$ $T^{\mathrm{loc}}$ and $T^{\mathrm{up}}$ denote
the local/global model data size in bits, the time required for generating
the synthetic data for all the devices, the time for the global model
broadcasting of the WFL server, the time for the local model training
of all the devices, and the time for the local model uploading of
all the devices, respectively. Then, the time of the third phase is
$N\left(T^{\mathrm{br}}+T^{\mathrm{loc}}+T^{\mathrm{up}}\right).$
We assume that there is a pre-determined maximum latency $T^{\mathrm{max}}$
for the whole WFL procedure, i.e.,
\begin{equation}
T^{\mathrm{syn}}+T^{\mathrm{down}}+N\left(T^{\mathrm{br}}+T^{\mathrm{loc}}+T^{\mathrm{up}}\right)\leq T^{\mathrm{max}}.
\end{equation}
Since the synthetic data are generated by the pre-trained AI generative
model in the AIGC server, we assume that $T^{\mathrm{syn}}$ is modeled
as a linear function of the amount of synthetic data generated for
all the devices \cite{xu2024unleashing}, i.e., $T^{\mathrm{syn}}=\varrho\sum_{k=1}^{K}D_{k}^{\mathrm{gen}}$.

According to the results in \cite{li2024filling}, the global learning
error (i.e., the global learning model accuracy) of the WFL depends
on the local training dataset sizes of the devices and the number
of global iterations, and can be modeled as
\begin{equation}
\triangle(\mathbf{D}^{\mathrm{gen}})=e^{\frac{N\left(\frac{\alpha}{K}\sum_{k=1}^{K}\left(D_{k}^{\mathrm{loc}}+D_{k}^{\mathrm{gen}}\right)^{-\beta}-\gamma-1\right)}{\zeta}},\label{eq:global-learning-error}
\end{equation}
where $\mathbf{D}^{\mathrm{gen}}=\{D_{k}^{\mathrm{gen}},k=1,\ldots,K\},$
$\zeta$ is a positive constant parameter, and $\alpha,$ $\beta,$
$\gamma$ are positive hyper-parameters that can be obtained through
curve fitting \cite{li2024filling}.

\subsection{Computation Model}

Let $f_{k},$ $w$ and $\tau$ denote the computing frequency of the
device $k,$ the number of CPU cycles of the devices to locally train
one data sample, and the local epoch, respectively. The computing
frequency of each device is capped by its maximum value, i.e., 
\begin{equation}
f_{k}\leq f_{k}^{\mathrm{max}},\forall k,
\end{equation}
where $f_{k}^{\mathrm{max}}$ is the maximum computing frequency of
the device $k.$ Then, the time of the device $k$ for a single-round
local model training is given by 
\begin{equation}
T_{k}^{\mathrm{loc}}=\frac{w\tau\left(D_{k}^{\mathrm{loc}}+D_{k}^{\mathrm{gen}}\right)}{f_{k}}.\label{eq:Tk_loc}
\end{equation}
Since each device has to finish the local model training within the
required time $T^{\mathrm{loc}},$ we have
\begin{equation}
T_{k}^{\mathrm{loc}}\leq T^{\mathrm{loc}},\forall k.
\end{equation}
In addition, the energy consumption per single-round local model training
of the device $k$ is written as
\begin{equation}
E_{k}^{\mathrm{loc}}=w\tau\varpi_{k}f_{k}^{2}\left(D_{k}^{\mathrm{loc}}+D_{k}^{\mathrm{gen}}\right),\label{eq:Ek_loc}
\end{equation}
where $\varpi_{k}$ denote the hardware energy coefficient \cite{xu2024latency}. 

\subsection{NOMA-Enhanced Communication Model}

Let $h_{k}$ and $g_{k}$ denote the channel gains from the BS to
the device $k,$ and from the device $k$ to the BS, respectively.
The system bandwidth is $B.$ There are three procedures that involve
wireless transmissions, i.e., synthetic data downloading from the
BS to the devices, global model broadcasting from the BS to the devices,
and local model uploading from the devices to the BS.

For synthetic data downloading, downlink NOMA is applied. Specifically,
according to the downlink NOMA principle, the messages intended for
the devices are transmitted simultaneously based on the superposition
coding. Each device can apply SIC to cancel the messages of other
devices whose channel gains are smaller than its own channel gain.
Without loss of the generality, we assume the devices are sorted in
the ascending order of the channel gain $h_{k},$ i.e., $h_{1}<h_{2}<\ldots<h_{K}$.
Let $p_{k}$ denote the transmit power of the message for the device
$k$. Thus, the achievable rate for synthetic data downloading of
the device $k$ is given by
\begin{equation}
R_{k}^{\mathrm{down}}=B\log_{2}\left(1+\frac{h_{k}p_{k}}{\sigma^{2}B+h_{k}\sum_{j>k}p_{j}}\right),\label{eq:rk}
\end{equation}
where $\sigma^{2}$ is the noise power spectral density. Since the
allocated synthetic data has to be finished transmitting within the
time $T^{\mathrm{down}},$ we have
\begin{equation}
T^{\mathrm{down}}R_{k}^{\mathrm{down}}\geq\Gamma D_{k}^{\mathrm{gen}},\forall k,
\end{equation}
where $\Gamma$ is the size of one data sample in bits. Furthermore,
the maximum transmit power of the BS is limited as
\begin{equation}
\sum_{k=1}^{K}p_{k}\leq P,
\end{equation}
where $P$ is the maximum transmit power of the BS.

For global model broadcasting, the BS is assumed to transmit at its
maximum power $P,$ and in order for all the devices to successfully
receive the global model, the broadcasting rate is chosen as the minimum
rate achievable for all the devices, i.e., $B\log_{2}\left(1+\frac{h_{1}P}{\sigma^{2}B}\right).$
Since the global model data has to be transmitted within the time
$T^{\mathrm{br}},$ we have
\begin{equation}
T^{\mathrm{br}}B\log_{2}\left(1+\frac{h_{1}P}{\sigma^{2}B}\right)\geq D^{\mathrm{mod}}.
\end{equation}

For local model uploading, uplink NOMA is adopted. Specifically, the
messages of all devices are transmitted to the BS simultaneously.
Let $q_{k}$ denote the transmit power of the device $k,$ and $Q_{k}$
denote the maximum transmit power of the device $k.$ Then, we have
\begin{equation}
q_{k}\leq Q_{k},\forall k.
\end{equation}
At the BS, the SIC is adopted to decode the messages of all the devices.
Denote by $\pi_{k}$ the SIC decoding order of the device $k$. Let
$\boldsymbol{\pi}=\{\pi_{k},\forall k\},$ and it belongs to the set
$\Pi$ of all possible SIC decoding orders of all $K$ messages. Thus,
the achievable rate of the device $k$ for local model uploading is
expressed as
\begin{align}
 & R_{k}^{\mathrm{up}}=B\log_{2}\left(1+\frac{g_{k}q_{k}}{\sigma^{2}B+\sum_{j=1,\pi_{j}>\pi_{k}}g_{j}q_{j}}\right).\label{eq:rate}
\end{align}
In order to upload the local model to the BS within the time $T^{\mathrm{up}},$
we have
\begin{equation}
T^{\mathrm{up}}R_{k}^{\mathrm{up}}\geq D^{\mathrm{mod}},\forall k.
\end{equation}
In addition, the energy consumption of the device $k$ for local model
uploading per single-round training is given by
\begin{equation}
E_{k}^{\mathrm{up}}=T^{\mathrm{up}}q_{k}.
\end{equation}

\section{NOMA+AIGC-Enhanced WFL \label{sec:Proposed-Algorithm-forall}}

\subsection{Problem Formulation}

Based on the system description in the previous section, the energy
consumption of the device $k$ for a single-round training is written
as
\begin{equation}
E_{k}=E_{k}^{\mathrm{loc}}+E_{k}^{\mathrm{up}}.\label{eq:Ek}
\end{equation}
We assume that there is an energy budget for each device, i.e., 
\begin{equation}
E_{k}\leq E_{k}^{\mathrm{max}},\forall k,
\end{equation}
where $E_{k}^{\mathrm{max}}$ is the maximum energy consumption of
the device $k$ per single-round training. 

Our focus is maximizing the learning performance under the constraints
analyzed in the system description. Specifically, the optimization
objective is minimizing the global learning error (i.e., to maximize
the global learning model accuracy), and the optimization variables
are the synthetic data allocation $\mathbf{D}^{\mathrm{gen}},$ the
time allocation $\mathbf{T}=\{T^{\mathrm{down}},T^{\mathrm{br}},T^{\mathrm{loc}},T^{\mathrm{up}}\},$
the transmit power allocation of the BS $\mathbf{p}=\{p_{k},k=1,\ldots,K\},$
the transmit power of the devices $\mathbf{q}=\{q_{k},k=1,\ldots,K\},$
the SIC decoding order $\boldsymbol{\pi},$ and the computing frequency
allocation $\mathbf{f}=\{f_{k},k=1,\ldots,K\}$. Mathematically, the
optimization problem for the NOMA+AIGC-enhanced WFL is formulated
as
\begin{subequations}
\label{eq:p1}
\begin{alignat}{1}
\min & \:\triangle(\mathbf{D}^{\mathrm{gen}})=e^{\frac{N\left(\frac{\alpha}{K}\sum_{k=1}^{K}\left(D_{k}^{\mathrm{loc}}+D_{k}^{\mathrm{gen}}\right)^{-\beta}-\gamma-1\right)}{\zeta}}\label{eq:p1-o}\\
\mathrm{s.t.} & \:\sum_{k=1}^{K}D_{k}^{\mathrm{gen}}\leq D^{\mathrm{gen}},\label{eq:p1-c1}\\
 & \:T^{\mathrm{syn}}+T^{\mathrm{down}}+N\left(T^{\mathrm{br}}+T^{\mathrm{loc}}+T^{\mathrm{up}}\right)\leq T^{\mathrm{max}},\label{eq:p1-c2}\\
 & \:f_{k}\leq f_{k}^{\mathrm{max}},\forall k,\label{eq:p1-c3}\\
 & \:T_{k}^{\mathrm{loc}}\leq T^{\mathrm{loc}},\forall k\label{eq:p1-c4}\\
 & \:T^{\mathrm{down}}R_{k}^{\mathrm{down}}\geq\Gamma D_{k}^{\mathrm{gen}},\forall k,\label{eq:p1-c6}\\
 & \:\sum_{k=1}^{K}p_{k}\leq P,\label{eq:p1-c7}\\
 & \:T^{\mathrm{br}}B\log_{2}\left(1+\frac{h_{1}P}{\sigma^{2}B}\right)\geq D^{\mathrm{mod}},\label{eq:p1-c8}\\
 & \:q_{k}\leq Q_{k},\forall k,\label{eq:p1-c9}\\
 & \:T^{\mathrm{up}}R_{k}^{\mathrm{up}}\geq D^{\mathrm{mod}},\forall k,\label{eq:p1-c10}\\
 & \:E_{k}\leq E_{k}^{\mathrm{max}},\forall k,\label{eq:p1-c11}\\
 & \:\boldsymbol{\pi}\in\Pi,\label{eq:p1-c12}\\
 & \:D_{k}^{\mathrm{gen}}\geq0,\forall k,\label{eq:p1-c13}\\
 & \:p_{k}\geq0,\forall k,\label{eq:p1-c14}\\
 & \:q_{k}\geq0,\forall k,\label{eq:p1-c15}\\
 & \:T^{\mathrm{down}}\geq0,T^{\mathrm{br}}\geq0,T^{\mathrm{loc}}\geq0,T^{\mathrm{up}}\geq0,\label{eq:p1-c16}\\
 & \:f_{k}\geq0,\forall k,\label{eq:p1-c17}\\
\mathrm{o.v.} & \:\mathbf{D}^{\mathrm{gen}},\mathbf{T},\mathbf{p},\mathbf{q},\boldsymbol{\pi},\mathbf{f}.
\end{alignat}
\end{subequations}
where the abbreviation \textquoteleft $\mathrm{s.t.}$\textquoteright{}
stands for \textquoteleft subject to\textquoteright{} and the abbreviation
\textquoteleft $\mathrm{o.v.}$\textquoteright{} stands for \textquoteleft optimization
variables\textquoteright . The constraint \eqref{eq:p1-c1} restricts
that the total transmitted synthetic data is smaller than the totally
synthetic data that can be generated in the AIGC server. The constraint
\eqref{eq:p1-c2} requires that the total time for $N$ WFL rounds
is less than the pre-defined maximum latency. The constraint \eqref{eq:p1-c3}
restricts the maximum computing frequency of each device. The constraint
\eqref{eq:p1-c4} requires that each device finishes the local model
training within the required time. The constraint \eqref{eq:p1-c6}
requires that the synthetic data can finish transmitting within the
synthetic data transmission time. The constraint \eqref{eq:p1-c7}
restricts the transmit power of the BS. The constraint \eqref{eq:p1-c8}
requires that the global model data can finish transmitting within
the global model broadcasting time. The constraint \eqref{eq:p1-c9}
restricts the transmit power of each device. The constraint \eqref{eq:p1-c10}
requires that the local model data can finish transmitting within
the local model uploading time. The constraint \eqref{eq:p1-c11}
restricts the energy consumption of each device. The constraint \eqref{eq:p1-c12}
restricts the SIC decoding order. The constraints \eqref{eq:p1-c13}-\eqref{eq:p1-c17}
restrict that the optimization variables are non-negative.

\subsection{Proposed Solution Framework}

Since $\boldsymbol{\pi}$ is discrete and the constraints \eqref{eq:p1-c6},
\eqref{eq:p1-c10}, \eqref{eq:p1-c11} are non-convex, the problem
\eqref{eq:p1} is a mixed integer nonlinear programming (MINLP) problem,
which is NP-hard in general, and its optimal solution is thus generally
intractable to find. Before proposing an algorithm to solve the problem
\eqref{eq:p1}, we present analysis to equivalently simplify the problem
\eqref{eq:p1}.
\begin{lem}
The constraint \eqref{eq:p1-c4} is satisfied with strict equality
by the optimal solution to the problem \eqref{eq:p1}.
\end{lem}
\begin{IEEEproof}
See Appendix \ref{subsec:Proof-of-Lemma1}.
\end{IEEEproof}
Lemma 1 indicates that each device shall use up the allocated local
model training time for saving energy consumption. From Lemma 1, the
optimal $\mathbf{f}$ is derived as
\begin{equation}
f_{k}=\frac{w\tau\left(D_{k}^{\mathrm{loc}}+D_{k}^{\mathrm{gen}}\right)}{T^{\mathrm{loc}}}.\label{eq:f}
\end{equation}
By inserting \eqref{eq:f} into the constraints \eqref{eq:p1-c3}
and \eqref{eq:p1-c11}, we have
\begin{align}
 & \frac{w\tau\left(D_{k}^{\mathrm{loc}}+D_{k}^{\mathrm{gen}}\right)}{T^{\mathrm{loc}}}\leq f_{k}^{\mathrm{max}},\forall k,\label{eq:new-p1-c3}\\
 & \frac{\varpi_{k}w^{3}\tau^{3}\left(D_{k}^{\mathrm{loc}}+D_{k}^{\mathrm{gen}}\right)^{3}}{(T^{\mathrm{loc}})^{2}}+E_{k}^{\mathrm{up}}\leq E_{k}^{\mathrm{max}},\forall k.\label{eq:new-p1-c11}
\end{align}
In addition, from \eqref{eq:global-learning-error}, it can be shown
that the objective function $\triangle(\mathbf{D}^{\mathrm{gen}})$
is a monotonically increasing function of $\sum_{k=1}^{K}\left(D_{k}^{\mathrm{loc}}+D_{k}^{\mathrm{gen}}\right)^{-\beta}$.
Thus, minimizing $\triangle(\mathbf{D}^{\mathrm{gen}})$ is equivalent
to minimizing $\sum_{k=1}^{K}\left(D_{k}^{\mathrm{loc}}+D_{k}^{\mathrm{gen}}\right)^{-\beta}$. 

Thanks to Lemma 1, the number of optimization variables is reduced.
After the above equivalent problem transformation, the problem \eqref{eq:p1}
is rewritten as
\begin{subequations}
\label{eq:p2}
\begin{alignat}{1}
\min & \:\sum_{k=1}^{K}\left(D_{k}^{\mathrm{loc}}+D_{k}^{\mathrm{gen}}\right)^{-\beta}\label{eq:p2-o}\\
\mathrm{s.t.} & \:\eqref{eq:p1-c1},\eqref{eq:p1-c2},\eqref{eq:p1-c6}-\eqref{eq:p1-c10},\eqref{eq:p1-c12}-\eqref{eq:p1-c16},\eqref{eq:new-p1-c3},\eqref{eq:new-p1-c11},\nonumber \\
\mathrm{o.v.} & \:\mathbf{D}^{\mathrm{gen}},\mathbf{T},\mathbf{p},\mathbf{q},\boldsymbol{\pi}.
\end{alignat}
\end{subequations}

\begin{figure}[!t]
\centering\includegraphics[width=0.7\columnwidth]{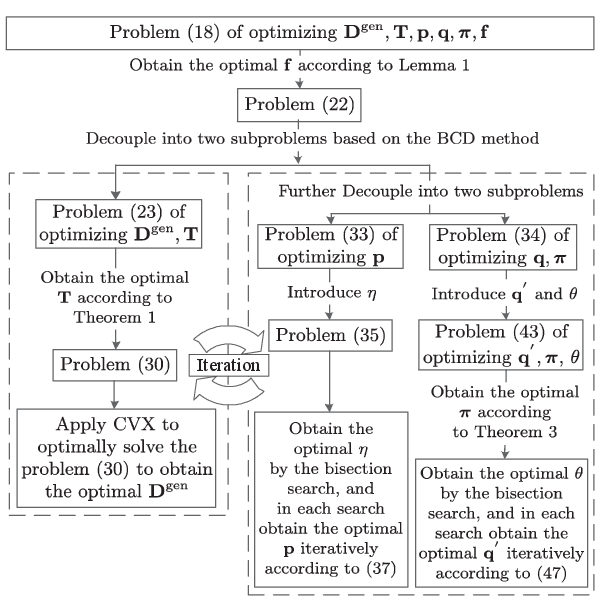}

\caption{Overall flowchart of solving the problem \eqref{eq:p1}.\label{fig:flowchart}}
\end{figure}

The problem \eqref{eq:p2} is still a MINLP problem, and its optimal
solution is hard to obtain. Therefore, we develop an efficient low-complexity
algorithm based on the BCD method to achieve a local optimal solution.
The overall flowchart of solving the problem \eqref{eq:p1} is shown
in Fig. \ref{fig:flowchart}. Specifically, the problem \eqref{eq:p2}
is decoupled into two subproblems. One subproblem on the left-hand-side
optimizes $\mathbf{D}^{\mathrm{gen}},\mathbf{T}$ with given $\mathbf{p},\mathbf{q},\boldsymbol{\pi}$
as shown in the left part of Fig. \ref{fig:flowchart}, and the other
one on the right-hand-side optimizes $\mathbf{p},\mathbf{q},\boldsymbol{\pi}$
with given $\mathbf{D}^{\mathrm{gen}},\mathbf{T}$ as shown in the
right part of Fig. \ref{fig:flowchart}. The two subproblems are solved
iteratively until the objective function value in \eqref{eq:p2-o}
converges. We will optimally solve the two subproblems in the next
two sections.

\section{Efficient Algorithm for Optimally Optimizing $\mathbf{D}^{\mathrm{gen}},\mathbf{T}$\label{sec:Proposed-Algorithm-D}}

In this subsection, the subproblem of optimizing $\mathbf{D}^{\mathrm{gen}},\mathbf{T}$
with given $\mathbf{p},\mathbf{q},\boldsymbol{\pi}$ (the left part
of Fig. \ref{fig:flowchart}) is investigated as
\begin{subequations}
\label{eq:p3}
\begin{alignat}{1}
\min & \:\sum_{k=1}^{K}\left(D_{k}^{\mathrm{loc}}+D_{k}^{\mathrm{gen}}\right)^{-\beta}\label{eq:p3-o}\\
\mathrm{s.t.} & \:\eqref{eq:p1-c1},\eqref{eq:p1-c2},\eqref{eq:p1-c6},\eqref{eq:p1-c8},\eqref{eq:p1-c10},\eqref{eq:p1-c13},\eqref{eq:p1-c16},\eqref{eq:new-p1-c3},\eqref{eq:new-p1-c11},\nonumber \\
\mathrm{o.v.} & \:\mathbf{D}^{\mathrm{gen}},\mathbf{T}.
\end{alignat}
\end{subequations}
In what follows, we present important properties of the problem \eqref{eq:p3}.
\begin{lem}
The optimal solution to the problem \eqref{eq:p3} satisfies the constraint
\eqref{eq:p1-c2} with strict equality.
\end{lem}
\begin{IEEEproof}
See Appendix \ref{subsec:Proof-of-Lemma2}.
\end{IEEEproof}
Lemma 2 indicates that the total allocated time for the whole WFL
procedure shall use up the maximum latency for saving the energy consumption
of the devices. From Lemma 2, we get
\begin{equation}
T^{\mathrm{syn}}+T^{\mathrm{down}}+N\left(T^{\mathrm{br}}+T^{\mathrm{loc}}+T^{\mathrm{up}}\right)=T^{\mathrm{max}}.\label{eq:T_all}
\end{equation}

\begin{lem}
The optimal solution to the problem \eqref{eq:p3} satisfies the constraint
\eqref{eq:p1-c6} with strict equality for a given $j=\arg\max_{k}\frac{D_{k}^{\mathrm{gen}}}{R_{k}^{\mathrm{down}}}$.
\end{lem}
\begin{IEEEproof}
See Appendix \ref{subsec:Proof-of-Lemma3}.
\end{IEEEproof}
Lemma 3 indicates that the synthetic data transmission time is determined
by the device with the maximum $\frac{D_{k}^{\mathrm{gen}}}{R_{k}^{\mathrm{down}}}.$
\begin{lem}
The optimal solution to the problem \eqref{eq:p3} satisfies the constraint
\eqref{eq:p1-c8} with strict equality.
\end{lem}
\begin{IEEEproof}
It can be proved similar to  Lemma 3, and is thus omitted here for
brevity.
\end{IEEEproof}
Lemma 4 indicates that the BS shall use up the allocated global model
downloading time for saving the energy consumption of the devices. 
\begin{lem}
The optimal solution to the problem \eqref{eq:p3} satisfies the constraint
\eqref{eq:p1-c10} with strict equality for a given $j=\min_{k}R_{k}^{\mathrm{up}}.$ 
\end{lem}
\begin{IEEEproof}
It can be proved similar to Lemma 3, and is thus omitted here for
brevity.
\end{IEEEproof}
Lemma 5 indicates that only the device with the minimum local model
uploading rate uses up the allocated local model uploading time, while
the other devices with higher local model uploading rate shall wait
until the end of the local model uploading time even though they finish
the local model uploading early. 

\begin{Thm}\label{thm:1}The optimal $\mathbf{T}$ for the problem
\eqref{eq:p3} given $\mathbf{D}^{\mathrm{gen}}$ is given by
\begin{align}
T^{\mathrm{down}}= & \Gamma\max_{k}\frac{D_{k}^{\mathrm{gen}}}{R_{k}^{\mathrm{down}}},\label{eq:T_down}\\
T^{\mathrm{br}}= & \frac{D^{\mathrm{mod}}}{B\log_{2}\left(1+\frac{h_{1}P}{\sigma^{2}B}\right)},\label{eq:T_br}\\
T^{\mathrm{up}}= & \frac{D^{\mathrm{mod}}}{\min_{k}R_{k}^{\mathrm{up}}},\label{eq:T_up}\\
T^{\mathrm{loc}}= & \check{T}^{\mathrm{loc}}-\frac{\varrho\sum_{k=1}^{K}D_{k}^{\mathrm{gen}}}{N}-\frac{\Gamma}{N}\max_{k}\frac{D_{k}^{\mathrm{gen}}}{R_{k}^{\mathrm{down}}},\label{eq:T_loc}
\end{align}
where
\begin{equation}
\check{T}^{\mathrm{loc}}=\frac{T^{\mathrm{max}}}{N}-\frac{D^{\mathrm{mod}}}{\min_{k}R_{k}^{\mathrm{up}}}-\frac{D^{\mathrm{mod}}}{B\log_{2}\left(1+\frac{h_{1}P}{\sigma^{2}B}\right)}.
\end{equation}
 \end{Thm}
\begin{IEEEproof}
The optimal $\mathbf{T}$ can be easily derived based on Lemmas 2-5.
Thus the details are omitted here for brevity. 
\end{IEEEproof}
Using Theorem \ref{thm:1}, the problem \eqref{eq:p3} is simplified
as
\begin{subequations}
\label{eq:p4}
\begin{alignat}{1}
\min_{\mathbf{D}^{\mathrm{gen}}} & \:\sum_{k=1}^{K}\left(D_{k}^{\mathrm{loc}}+D_{k}^{\mathrm{gen}}\right)^{-\beta}\label{eq:p4-o}\\
\mathrm{s.t.} & \:D_{k}^{\mathrm{gen}}+\frac{\Gamma f_{k}^{\mathrm{max}}}{Nw\tau}\max_{j}\frac{D_{j}^{\mathrm{gen}}}{R_{j}^{\mathrm{down}}}+\frac{f_{k}^{\mathrm{max}}\varrho\sum_{k=1}^{K}D_{k}^{\mathrm{gen}}}{Nw\tau}\nonumber \\
 & \:\leq\frac{f_{k}^{\mathrm{max}}\check{T}^{\mathrm{loc}}}{w\tau}-D_{k}^{\mathrm{loc}},\forall k,\label{eq:p4-c1}\\
 & \:\sqrt{\!\frac{\varpi_{k}w^{3}\tau^{3}\left(D_{k}^{\mathrm{loc}}+D_{k}^{\mathrm{gen}}\right)^{3}}{E_{k}^{\mathrm{max}}-E_{k}^{\mathrm{up}}}}+\frac{\Gamma}{N}\max_{j}\frac{D_{j}^{\mathrm{gen}}}{R_{j}^{\mathrm{down}}}\nonumber \\
 & \:+\frac{\varrho\sum_{k=1}^{K}D_{k}^{\mathrm{gen}}}{N}\leq\!\check{T}^{\mathrm{loc}},\forall k,\label{eq:p4-c2}\\
 & \:\eqref{eq:p1-c1},\eqref{eq:p1-c13}.\nonumber 
\end{alignat}
\end{subequations}
It can be verified that the objective function in \eqref{eq:p4-o}
is a convex function of $\mathbf{D}^{\mathrm{gen}}$ and all the constraints
are convex or linear with respect to $\mathbf{D}^{\mathrm{gen}}$.
Thus, the problem \eqref{eq:p4} is a convex optimization problem
and can be optimally solved via CVX \cite{cvx}.

The proposed algorithm to optimally optimize $\mathbf{D}^{\mathrm{gen}},\mathbf{T}$
is summarized in Algorithm \ref{alg:1}. Since the complexity of solving
the problem \eqref{eq:p4} is $\mathcal{O}(K^{3})$ \cite{yang2022sum},
the complexity of Algorithm \ref{alg:1} is $\mathcal{O}(K^{3}).$

\begin{algorithm}[!t]
\begin{algorithmic}[1]
\STATE Solve the problem \eqref{eq:p4} using CVX to obtain $\mathbf{D}^{\mathrm{gen}}$.
\STATE Obtain the closed-form expressions for $T^{\mathrm{down}},$ $T^{\mathrm{br}},$ $T^{\mathrm{up}},$ and  $T^{\mathrm{loc}}$ from \eqref{eq:T_down}, \eqref{eq:T_br}, \eqref{eq:T_up}, and \eqref{eq:T_loc}, respectively.
\end{algorithmic}\caption{Proposed algorithm to optimally optimize $\mathbf{D}^{\mathrm{gen}},\mathbf{T}$
based on the convex optimization.\label{alg:1}}
\end{algorithm}

\section{Efficient Algorithm for Optimally Optimizing $\mathbf{p},\mathbf{q},\boldsymbol{\pi}$\label{sec:Proposed-Algorithm-p}}

In this section, the subproblem of optimizing $\mathbf{p},\mathbf{q},\boldsymbol{\pi}$
with given $\mathbf{D}^{\mathrm{gen}},\mathbf{T}$ (the right part
of Fig. \ref{fig:flowchart}) is investigated. It can be shown that
the constraints related to $\mathbf{p}$ and the constraints related
to $\mathbf{q},\boldsymbol{\pi}$ are different. Therefore, the subproblem
of optimizing $\mathbf{p},\mathbf{q},\boldsymbol{\pi}$ can be decoupled
into two problems, given by
\begin{alignat}{1}
\mathrm{Find} & \:\mathbf{p}\label{eq:p10}\\
\mathrm{s.t.} & \:\eqref{eq:p1-c6},\eqref{eq:p1-c7},\eqref{eq:p1-c14},\nonumber 
\end{alignat}
and
\begin{alignat}{1}
\mathrm{Find} & \:\mathbf{q},\boldsymbol{\pi}\label{eq:p11}\\
\mathrm{s.t.} & \:\eqref{eq:p1-c9},\eqref{eq:p1-c10},\eqref{eq:p1-c12},\eqref{eq:p1-c15},\eqref{eq:new-p1-c11}.\nonumber 
\end{alignat}
Both the problems in \eqref{eq:p10} and \eqref{eq:p11} try to find
feasible solutions, since the objective function in the original problem
\eqref{eq:p2} does not depend on the optimization variables $\mathbf{p},\mathbf{q},\boldsymbol{\pi}$.
However, in order to find feasible solutions which can lead to more
favorable results such that the subproblem \eqref{eq:p3} has higher
objective function value, the objective functions in the problems
\eqref{eq:p10} and \eqref{eq:p11} are modified respectively as
\begin{alignat}{1}
\max_{\mathbf{p}} & \:\min_{k}\frac{R_{k}^{\mathrm{down}}}{D_{k}^{\mathrm{gen}}}\label{eq:p12}\\
\mathrm{s.t.} & \:\eqref{eq:p1-c6},\eqref{eq:p1-c7},\eqref{eq:p1-c14},\nonumber 
\end{alignat}
and
\begin{alignat}{1}
\max_{\mathbf{q},\boldsymbol{\pi}} & \:\min_{k}R_{k}^{\mathrm{up}}\label{eq:p13}\\
\mathrm{s.t.} & \:\eqref{eq:p1-c9},\eqref{eq:p1-c10},\eqref{eq:p1-c12},\eqref{eq:p1-c15},\eqref{eq:new-p1-c11}.\nonumber 
\end{alignat}
It is shown that by selecting the objective function as expressed
in \eqref{eq:p12}, the constraint \eqref{eq:p1-c6} can be more relaxed
in the subproblem \eqref{eq:p3}, which can lead to higher objective
function value. It is also shown that by selecting the objective function
as expressed in \eqref{eq:p13}, the constraint \eqref{eq:p1-c10}
can be more relaxed in the subproblem \eqref{eq:p3}, which can also
lead to higher objective function value. In what follows, we first
solve the problem \eqref{eq:p12}, and then solve the problem \eqref{eq:p13}.

\subsection{Proposed algorithm for optimally optimizing $\mathbf{p}$\label{subsec:Proposed-algorithm-for-P}}

By introducing an auxiliary variable $\eta=\min_{k}\frac{R_{k}^{\mathrm{down}}}{D_{k}^{\mathrm{gen}}}$,
the problem \eqref{eq:p12} can be reformulated as
\begin{subequations}
\label{eq:p14}
\begin{alignat}{1}
\max_{\mathbf{p},\eta} & \:\eta\label{eq:p14-o}\\
\mathrm{s.t.} & \:R_{k}^{\mathrm{down}}\geq\eta D_{k}^{\mathrm{gen}},\forall k,\label{eq:p14-c1}\\
 & \:\eqref{eq:p1-c6},\eqref{eq:p1-c7},\eqref{eq:p1-c14}.\nonumber 
\end{alignat}
\end{subequations}
It is noted that the constraint \eqref{eq:p1-c6} is inactive as long
as the problem is feasible, i.e., $\eta\geq\frac{\Gamma}{T^{\mathrm{down}}}.$
Next, we present an important property of the optimal solution to
the problem \eqref{eq:p14}.
\begin{lem}
The optimal solution to the problem \eqref{eq:p14} satisfies the
constraint \eqref{eq:p14-c1} with strict equality. 
\end{lem}
\begin{IEEEproof}
See Appendix \ref{subsec:Proof-of-Lemma6}
\end{IEEEproof}
From Lemma 6, the optimal $\mathbf{p}$ given $\eta$ satisfies the
following equalities
\begin{equation}
B\log_{2}\left(1+\frac{h_{k}p_{k}}{\sigma^{2}B+h_{k}\sum_{j>k}p_{j}}\right)=\eta D_{k}^{\mathrm{gen}},\forall k,\label{eq:sum_pk}
\end{equation}

\begin{Thm}\label{thm:2}The optimal $p_{k},k=1,\ldots,K$ for the
problem \eqref{eq:p14} given $\eta$ can be sequentially determined
from $p_{K}$ to $p_{1}$ according to
\begin{equation}
p_{k}=\left(2^{\frac{\eta D_{k}^{\mathrm{gen}}}{B}}-1\right)\left(\frac{\sigma^{2}B}{h_{k}}+\sum_{j>k}p_{j}\right),\forall k.\label{eq:pk}
\end{equation}
\end{Thm}
\begin{IEEEproof}
This theorem is a direct result from Lemma 6. By rewriting \eqref{eq:sum_pk},
we can have \eqref{eq:pk}, where the optimal $p_{k}$ is determined
by $p_{j},j>k$. This completes the proof. 
\end{IEEEproof}
After obtaining the optimal $\mathbf{p}$, what remains is optimizing
$\eta$ in the problem \eqref{eq:p14}. From \eqref{eq:pk}, it shows
that $p_{k}$ is an increasing function of $\eta.$ Therefore, if
$\eta$ is larger than the optimal $\eta,$ the constraint \eqref{eq:p1-c7}
will be violated. Accordingly, the optimal $\eta$ for the problem
\eqref{eq:p14} can be derived by a bisection search of $\eta$, where
$\eta$ is increased if the constraint \eqref{eq:p1-c7} with given
$\eta$ is obeyed, and is decreased if the constraint \eqref{eq:p1-c7}
with given $\eta$ is violated.

The proposed algorithm to optimize $\mathbf{p}$ is summarized in
Algorithm \ref{alg:2}. Since the bisection search method converges
in finite number of iterations which is independent of the number
of devices \cite{convexop2004}, the complexity of Algorithm \ref{alg:2}
is merely linear in the number of devices, i.e., $\mathcal{O}(K).$

\begin{algorithm}[!t]
\begin{algorithmic}[1]
\STATE Initialize $\eta_{min}$ and $\eta_{max}$.
\REPEAT
\STATE $\eta=\frac{\eta_{min}+\eta_{max}}{2}.$
\FORALL{$k=K$ to $1$}
\STATE Obtain $p_{k}$ from \eqref{eq:pk}.
\ENDFOR
\IF{the constraint \eqref{eq:p1-c7} is obeyed}
\STATE $\eta_{min}=\eta.$
\ELSE
\STATE $\eta_{max}=\eta.$
\ENDIF
\UNTIL{$\eta$ converges.}

\end{algorithmic}\caption{Proposed algorithm to obtain the optimal $\mathbf{p}$.\label{alg:2}}
\end{algorithm}

\subsection{Proposed algorithm for optimally optimizing $\mathbf{q},\boldsymbol{\pi}$\label{subsec:Proposed-algorithm-for-Q}}

By integrating the constraints \eqref{eq:p1-c9}, \eqref{eq:p1-c15}
and \eqref{eq:new-p1-c11}, it follows that
\begin{equation}
0\leq q_{k}\leq q_{k}^{\mathrm{max}},\forall k,\label{eq:p15-c1}
\end{equation}
where
\begin{equation}
q_{k}^{\mathrm{max}}=\min\left(Q_{k},\frac{E_{k}^{\mathrm{max}}}{T^{\mathrm{up}}}-\frac{\varpi_{k}w^{3}\tau^{3}\left(D_{k}^{\mathrm{loc}}+D_{k}^{\mathrm{gen}}\right)^{3}}{(T^{\mathrm{loc}})^{2}T^{\mathrm{up}}}\right).
\end{equation}
Furthermore, it shows that the constraint \eqref{eq:p1-c10} is redundant
given the objective function formulated in \eqref{eq:p13}, as long
as the problem \eqref{eq:p13} is feasible. Thus, the problem \eqref{eq:p13}
is rewritten as
\begin{alignat}{1}
\max_{\mathbf{q},\boldsymbol{\pi}} & \:\min_{k}R_{k}^{\mathrm{up}}\label{eq:p15}\\
\mathrm{s.t.} & \:\eqref{eq:p1-c12},\eqref{eq:p15-c1}.\nonumber 
\end{alignat}
By defining $\mathbf{q}^{'}=\{q_{k}^{'},k=1,\ldots,K\}$, where $q_{k}^{'}=\frac{q_{k}}{q_{k}^{\mathrm{max}}}$,
the problem \eqref{eq:p15} can be reformulated as
\begin{subequations}
\label{eq:p16}
\begin{alignat}{1}
\max_{\mathbf{q}^{'},\boldsymbol{\pi}} & \:\min_{k}R_{k}^{\mathrm{up}}\label{eq:p16-o}\\
\mathrm{s.t.} & \:0\leq q_{k}^{'}\leq1,\forall k,\label{eq:p16-c1}\\
 & \:\eqref{eq:p1-c12},\nonumber 
\end{alignat}
\end{subequations}
where $R_{k}^{\mathrm{up}}$ is rewritten as
\begin{equation}
R_{k}^{\mathrm{up}}=B\log_{2}\left(1+\frac{g_{k}q_{k}^{\mathrm{max}}q_{k}^{'}}{\sigma^{2}B+\sum_{j=1,\pi_{j}>\pi_{k}}g_{j}q_{j}^{\mathrm{max}}q_{j}^{'}}\right).
\end{equation}
By introducing an auxiliary variable $\theta=\min_{k}R_{k}^{\mathrm{up}}$,
the problem \eqref{eq:p16} can be reformulated as
\begin{subequations}
\label{eq:p17}
\begin{alignat}{1}
\max_{\mathbf{q}^{'},\boldsymbol{\pi},\theta} & \:\theta\label{eq:p17-o}\\
\mathrm{s.t.} & \:R_{k}^{\mathrm{up}}\geq\theta,\forall k,\label{eq:p17-c1}\\
 & \:\eqref{eq:p1-c12},\eqref{eq:p16-c1}.\nonumber 
\end{alignat}
\end{subequations}

Due to the discrete constraint \eqref{eq:p1-c12}, it is intractable
to find the optimal solution to the problem \eqref{eq:p17-o} by standard
optimization methods. In what follows, we derive the optimal SIC decoding
order $\boldsymbol{\pi}$ by exploring the problem structure.

\begin{Thm}\label{thm:3}The optimal $\boldsymbol{\pi}$ for the
problem \eqref{eq:p17} is in the descending order of $g_{k}q_{k}^{\mathrm{max}}.$\end{Thm}
\begin{IEEEproof}
See Appendix \ref{subsec:Proof-ofTheorem1}.
\end{IEEEproof}
From Theorem \ref{thm:3}, the optimal decoding order $\boldsymbol{\pi}$
for the problem \eqref{eq:p17} can be obtained, and the problem \eqref{eq:p17}
is simplified as
\begin{alignat}{1}
\max_{\mathbf{q}^{'},\theta} & \:\theta\label{eq:p19}\\
\mathrm{s.t.} & \:\eqref{eq:p16-c1},\eqref{eq:p17-c1}.\nonumber 
\end{alignat}
Since the above problem given $\theta$ is feasible only when $\theta$
is smaller than or equal to the optimal $\theta$, the optimal $\theta$
can be obtained by a simple bisection search of $\theta$, where in
each search, the problem of optimizing $\mathbf{q}^{'}$ given $\theta$
is investigated as 
\begin{alignat}{1}
\textrm{find} & \:\mathbf{q}^{'}\label{eq:p20}\\
\mathrm{s.t.} & \:\eqref{eq:p16-c1},\eqref{eq:p17-c1}.\nonumber 
\end{alignat}

\begin{lem}
A feasible solution to the problem \eqref{eq:p20} satisfies the constraint
\eqref{eq:p17-c1} with strict equality. 
\end{lem}
\begin{IEEEproof}
See Appendix \ref{subsec:Proof-of-Lemma7}.
\end{IEEEproof}
From Lemma 7, the feasible solution to the problem \eqref{eq:p20}
satisfies the following equalities
\begin{align}
 & B\log_{2}\left(1+\frac{g_{k}q_{k}^{\mathrm{max}}q_{k}^{'}}{\sigma^{2}B+\sum_{j=1,\pi_{j}>\pi_{k}}g_{j}q_{j}^{\mathrm{max}}q_{j}^{'}}\right)=\theta,\forall k.\label{eq:qk-0}
\end{align}
\begin{Thm}\label{thm:4}The feasible $q_{k}^{'},k=1,\ldots,K$ for
the problem \eqref{eq:p20} can be recursively obtained from the last
decoding device to the first decoding device according to
\begin{equation}
q_{k}^{'}=\frac{2^{\frac{\theta}{B}}-1}{g_{k}q_{k}^{\mathrm{max}}}\left(\sigma^{2}B+\sum_{j=1,\pi_{j}>\pi_{k}}g_{j}q_{j}^{\mathrm{max}}q_{j}^{'}\right),\forall k.\label{eq:qk}
\end{equation}
\end{Thm}
\begin{IEEEproof}
This theorem is a direct result from Lemma 7. After rewritting \eqref{eq:qk-0},
we can have \eqref{eq:qk}, which shows that $q_{k}^{'}$ depends
only on $q_{j}^{'},\pi_{j}>\pi_{k}$. This completes the proof. 
\end{IEEEproof}
After $\mathbf{q}^{'}$ has been obtained from \eqref{eq:qk}, we
can check whether the constraint \eqref{eq:p16-c1} is satisfied by
the obtained $\mathbf{q}^{'}$. Specifically, the problem \eqref{eq:p20}
with the given $\theta$ is infeasible if the constraint \eqref{eq:p16-c1}
is not satisfied, and is feasible otherwise.

The proposed algorithm to optimally optimize $\mathbf{q},\boldsymbol{\pi}$
is summarized in Algorithm \ref{alg:3}. Since the optimal decoding
order is derived in closed form and the convergence of the bisection
search method does not depend on the number of devices \cite{convexop2004},
the complexity of Algorithm \ref{alg:3} is only linear in the number
of devices, i.e., $\mathcal{O}(K).$

\begin{algorithm}[!t]
\begin{algorithmic}[1]

\STATE Obtain the optimal decoding order $\boldsymbol{\pi}$ in the descending order of $g_{k}q_{k}^{\mathrm{max}}.$

\REPEAT
\STATE $\theta=\frac{\theta_{min}+\theta_{max}}{2}.$
\STATE Obtain $\mathbf{q}^{'}$ recursively  from the last decoding device to the first decoding device from \eqref{eq:qk}.

\IF{the constraint \eqref{eq:p16-c1} is obeyed}
\STATE $\theta_{min}=\theta.$
\ELSE
\STATE $\theta_{max}=\theta.$
\ENDIF
\UNTIL{$\theta$ converges.}
\STATE Obtain $q_{k}=q_{k}^{\mathrm{max}}q_{k}^{'},k=1,\ldots,K.$
\end{algorithmic}\caption{Proposed algorithm to obtain the optimal $\mathbf{q},\boldsymbol{\pi}$.\label{alg:3}}
\end{algorithm}

\subsection{Convergence and Complexity Analysis of the Overall Proposed Algorithm}

\begin{algorithm}[!t]
\begin{algorithmic}[1]
\STATE Initialize $\mathbf{D}^{\mathrm{gen}},\mathbf{T},\mathbf{p},\mathbf{q},\boldsymbol{\pi}$.
\REPEAT
\STATE Optimizes $\mathbf{D}^{\mathrm{gen}},\mathbf{T}$ with given $\mathbf{p},\mathbf{q},\boldsymbol{\pi}$ using Algorithm \ref{alg:1}.
\STATE Optimizes $\mathbf{p}$ with given $\mathbf{D}^{\mathrm{gen}},\mathbf{T}$ using Algorithm \ref{alg:2}.
\STATE Optimizes $\mathbf{q},\boldsymbol{\pi}$ with given $\mathbf{D}^{\mathrm{gen}},\mathbf{T}$ using Algorithm \ref{alg:3}.
\UNTIL{the objective function value in \eqref{eq:p2-o} converges.}
\STATE Obtain $\mathbf{f}$ from \eqref{eq:f}.
\end{algorithmic}\caption{Overall proposed algorithm for solving the problem \eqref{eq:p1}
based on the BCD method.\label{alg:4}}
\end{algorithm}

The overall proposed algorithm for solving the problem \eqref{eq:p1}
is summarized in Algorithm \ref{alg:4}. In this subsection, we provide
convergence and complexity analysis of the overall proposed algorithm. 

The convergence of the proposed algorithm is affected by the BCD method
used for iteratively solving the problem \eqref{eq:p2}, i.e., the
algorithm \ref{alg:1} for optimizing $\mathbf{D}^{\mathrm{gen}},\mathbf{T}$
and the algorithms \ref{alg:2} and \ref{alg:3} for optimizing $\mathbf{p},\mathbf{q},\boldsymbol{\pi}$
are iteratively performed until convergence. The following proposition
presents the convergence analysis of the adopted BCD method.

\begin{Prop}The BCD method in Algorithm \ref{alg:4} used for iteratively
solving the problem \eqref{eq:p2} converges to a local optimal solution
the problem \eqref{eq:p2}.\end{Prop}
\begin{IEEEproof}
See Appendix \ref{subsec:Proof-of-Proposition1}.
\end{IEEEproof}
The above Proposition has shown the convergence of the proposed algorithm
to a locally optimal solution. Then, we provide the complexity analysis.
Note that the BCD method converges in finite number of iterations
which is independent of the number of devices. Thus, based on the
complexity analysis in Section \ref{sec:Proposed-Algorithm-D}, Section
\ref{subsec:Proposed-algorithm-for-P} and Section \ref{subsec:Proposed-algorithm-for-Q},
the total complexity of Algorithm \ref{alg:4} is $\mathcal{O}(K^{3}+2K)$,
which is only polynomial in the number of devices.

\section{Simulation Results\label{sec:Simulation-Results}}

This section provides illustrative simulation results to demonstrate
the effectiveness of the proposed NOMA+AIGC-enhanced WFL scheme. Unless
otherwise noted, the simulation parameters are set as follows. The
setting of the AIGC server is similar to \cite{li2024filling}, where
eight RTX A5000 GPUs are equipped by the AIGC server and synthesizing
one data sample requires approximately $0.0646$ s, i.e., $\varrho=0.0646$.
The number of devices is $K=15$, where the devices are randomly distributed
around the BS within the distance range $[150,300]$ m. The wireless
channels are assumed to follow Rayleigh fading, where the channel
gains $h_{k}$ and $g_{k}$ are modeled as $h_{k}=\hat{h}_{k}\tilde{h}_{k}$
and $g_{k}=\hat{g}_{k}\tilde{g}_{k}$, respectively. Specifically,
the $\hat{h}_{k}$ (or $\hat{g}_{k}$) is the mean value of $h_{k}$
(or $g_{k}$) and is modeled as $128.1+37.6\log_{10}(d)$ in dB \cite{access2010further},
where $d$ is the distance in km. The $\tilde{h}_{k}$ (or $\tilde{g}_{k})$
is an exponentially distributed random variable with unit mean. The
$D_{k}^{\mathrm{loc}}$, $D^{\mathrm{gen}}$, $D^{\mathrm{mod}},$
$w,$ $f_{k}^{\mathrm{max}}$ are assumed to be uniformly distributed
within $[300,500]$ samples, $[3000,5000]$ samples, $[1.5,2.5]$
Mbits, $[1,2]\times10^{6}$ cycles, and $[1,2]$ GHz, respectively.
In addition, we set $B=1$ MHz, $\sigma^{2}=-160$ dBm/Hz, $N=100,$
$T^{\mathrm{max}}=900$ s, $\tau=1$, $\zeta=50,$ $\alpha=3.819,$
$\beta=0.198,$ $\gamma=0.231$ \cite{li2024filling}, $P=35$ dBm,
$Q_{k}=20$ dBm, $\varpi=10^{-27}$ \cite{bouzinis2022wireless},
$\Gamma=20$ Kbits, and $E_{k}^{max}=1.2$ Joule. 

For the purpose of comparison, the following four schemes developed
in existing literature or coined for benchmarking are considered:
\begin{itemize}
\item FDMA+AIGC: In this scheme, FDMA is used for both synthetic data downloading
and local model uploading as in \cite{li2024filling}. Note that since
\cite{li2024filling} did not consider the synthetic data downloading
phase, we modify the algorithm developed in \cite{li2024filling}
to suit our considered model.
\item TDMA+AIGC: In this scheme, TDMA is used for both synthetic data downloading
and local model uploading.
\item NOMA-w/o-AIGC: In this scheme, synthetic data are not transmitted
to the devices, and the proposed NOMA scheme is used for the local
model uploading.
\item FDMA-w/o-AIGC: In this scheme, synthetic data are not transmitted
to the devices, and FDMA is used for local model uploading as in \cite{li2024filling}.
\end{itemize}

\begin{figure}[!t]
\centering\includegraphics[width=0.42\columnwidth]{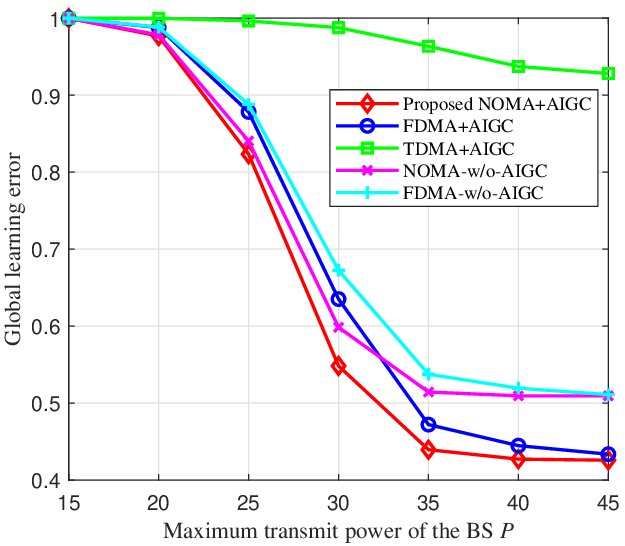}

\caption{Impact of $P$ on the learning performance.\label{fig:1}}
\end{figure}

Fig. \ref{fig:1} illustrates the impact of the maximum transmit power
of the BS $P$ on the learning performance. It shows that the learning
performance of the proposed NOMA+AIGC scheme outperforms all the other
schemes including FDMA+AIGC. This indicates that our proposed scheme
is more effective in improving the WFL performance. Meanwhile, it
shows that the learning performance with AIGC improves a lot compared
to that without AIGC. This is because by downloading synthetic data
from the server, more training data are available for local training
to improve the training performance. It is also shown that TDMA+AIGC
even underperforms the NOMA and the FDMA schemes without AIGC. This
is due to the fact that the devices that are unscheduled for synthetic
data downloading in TDMA shall wait and this significantly lowers
the time resource utilization and makes the maximum latency constraint
vulnerable. 

As $P$ increases, Fig. \ref{fig:1} shows that the learning performance
improves. The reason for this is that a higher $P$ can let the server
transmit more synthetic data to the devices to improve the learning
performance for the schemes with AIGC. Meanwhile, for the schemes
without AIGC, a higher $P$ can let the BS broadcast the global model
to the devices with less time, and thus it is easier to satisfy the
maximum latency constraint and make the investigated problem feasible,
since we set the learning error to one when the problem is infeasible.
When $P$ is very large, it shows that the learning performance saturates.
This is since the system will be restricted by the other factors such
as the energy consumption constraint and the latency constraint, and
no more synthetic data can be transmitted to the devices if $P$ is
very high. 

\begin{figure}[!t]
\centering\includegraphics[width=0.42\columnwidth]{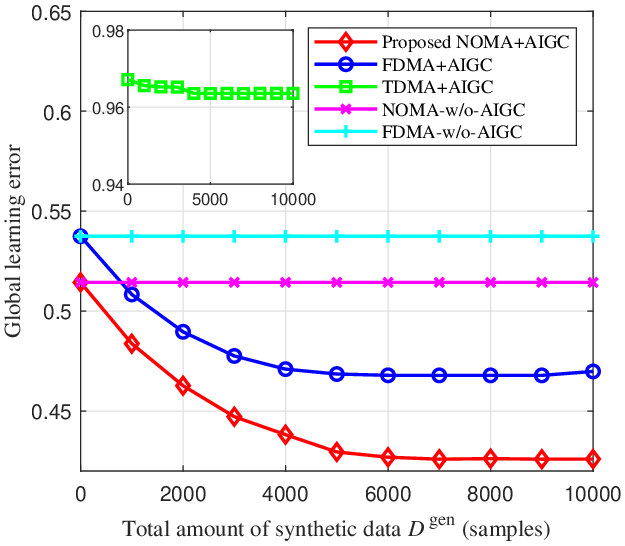}

\caption{Impact of $D^{\mathrm{gen}}$ on the learning performance.\label{fig:2}}
\end{figure}

Fig. \ref{fig:2} illustrates the impact of the total amount of synthetic
data $D^{\mathrm{gen}}$ on the learning performance. It shows that
the learning performance of the schemes without AIGC remains unchanged
as $D^{\mathrm{gen}}$ increases. This is since the schemes without
AIGC do not use synthetic data for local training. It also shows that
the schemes with AIGC achieve better learning performance as $D^{\mathrm{gen}}$
increases. This is because higher amount of synthetic data available
at the server can let the server transmit more synthetic data to the
devices. Whereas when $D^{\mathrm{gen}}$ is very large, a larger
$D^{\mathrm{gen}}$ cannot improve the learning performance further.
This is since the energy consumption constraint and the latency constraint
may restrict the learning performance and the synthetic data received
by the devices cannot be increased even if the server has more synthetic
data available. Furthermore, it shows that the learning performance
of the proposed NOMA+AIGC scheme is the best among all the schemes
and the performance improvement is more obvious when $D^{\mathrm{gen}}$
is larger. This means that the proposed scheme is more effective in
utilizing the synthetic data if more synthetic data are available
at the server for improving the learning performance.

\begin{figure}[!t]
\centering\includegraphics[width=0.42\columnwidth]{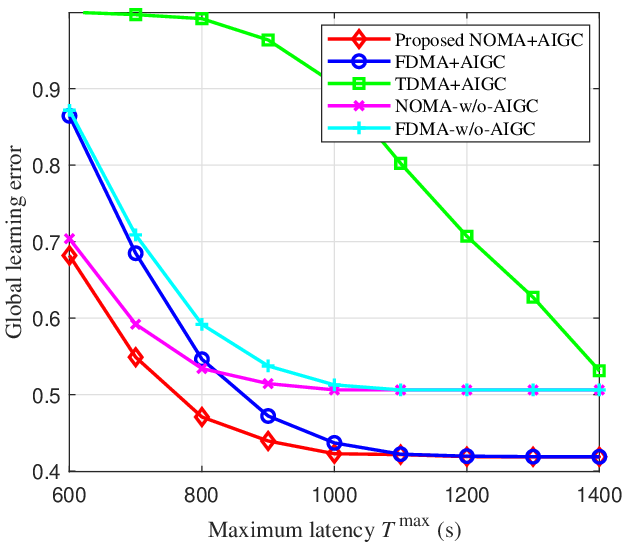}

\caption{Impact of $T^{\mathrm{max}}$ on the learning performance.\label{fig:3}}
\end{figure}

Fig. \ref{fig:3} illustrates the impact of the maximum latency constraint
$T^{\mathrm{max}}$ on the learning performance. It shows that all
the schemes can achieve better learning performance as $T^{\mathrm{max}}$
increases. This is because a larger value of $T^{\mathrm{max}}$ can
let the server transmit more synthetic data to the devices for better
learning performance of the schemes with AIGC, while the maximum latency
constraint is easier to be satisfied for improving the learning performance
of the schemes without AIGC. It also shows that when $T^{\mathrm{max}}$
is very large, the schemes with/without AIGC converge to the same
learning performance. This is since a very large $T^{\mathrm{max}}$
can let the server transmit as much data as possible under other constraints
such as the energy consumption constraint, and the efficiencies of
different multiple access schemes will have no impact on the learning
performance. There is still a gap between the schemes with and without
AIGC, since the schemes without AIGC are only easier to satisfy the
problem constraints when $T^{\mathrm{max}}$ is larger, and the learning
performance will remain the same if the problem is feasible. 

In addition, Fig. \ref{fig:3} shows that the performance improvement
of the proposed scheme compared to FDMA+AIGC is more obvious when
$T^{\mathrm{max}}$ is smaller. This is due to the fact that the proposed
scheme can utilize the time resource more efficiently than FDMA+AIGC.
It also shows that the performance improvement of TDMA+AIGC with the
increases of $T^{\mathrm{max}}$ is more obvious than the other schemes.
This is due to the fact that the time resource utilization efficiency
of TDMA+AIGC is low and a larger $T^{\mathrm{max}}$ can efficiently
offset the impact of low time resource utilization efficiency on the
learning performance.

\begin{figure}[!t]
\centering\includegraphics[width=0.42\columnwidth]{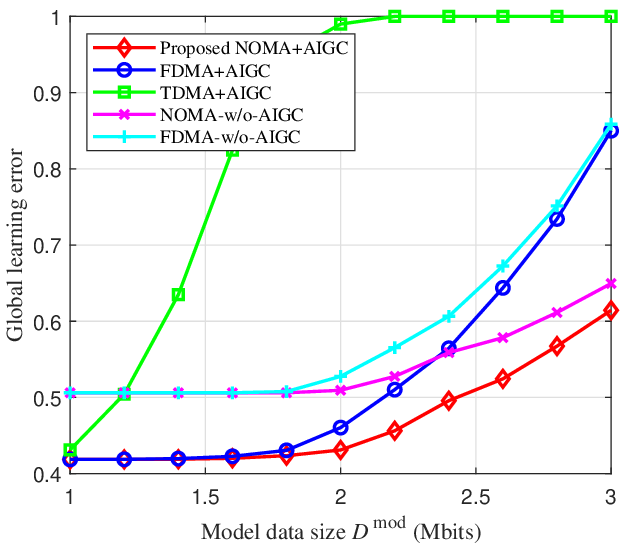}

\caption{Impact of $D^{\mathrm{mod}}$ on the learning performance.\label{fig:4}}
\end{figure}

Fig. \ref{fig:4} illustrates the impact of the model data size $D^{\mathrm{mod}}$
on the learning performance. It shows that the learning performance
degrades as $D^{\mathrm{mod}}$ increases. This is because a larger
$D^{\mathrm{mod}}$ requires more time for local model uploading and
global model broadcasting, which leads to less time for synthetic
data downloading and may also render the problem to be infeasible.
It also shows that TDMA+AIGC quickly saturates to the worst learning
performance. This is since TDMA+AIGC has the lowest time resource
utilization efficiency, and a larger $D^{\mathrm{mod}}$ will make
the required time for local model uploading and global model broadcasting
much larger. Moreover, it shows that the proposed NOMA+AIGC scheme
achieves the highest learning performance and the performance degradation
due to the increase of $D^{\mathrm{mod}}$ is the least among all
the schemes. This means that the proposed scheme can minimize the
impact of the increased time needed for local model uploading and
global model broadcasting to guarantee the learning performance. 

\begin{figure}[!t]
\centering\includegraphics[width=0.42\columnwidth]{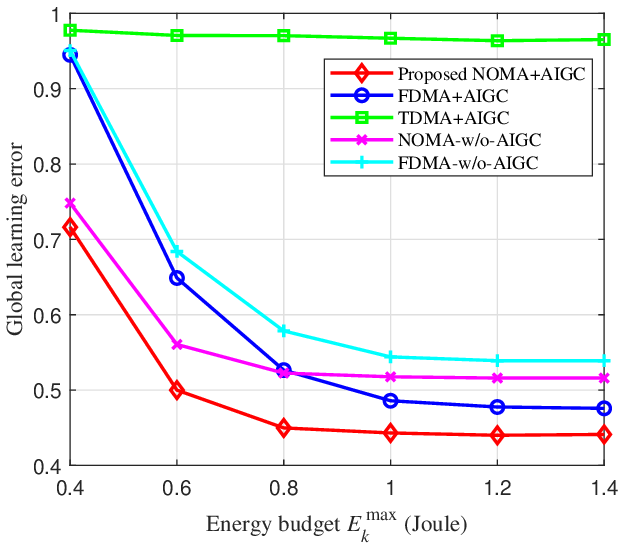}

\caption{Impact of $E_{k}^{\mathrm{max}}$ on the learning performance.\label{fig:5}}
\end{figure}

Fig. \ref{fig:5} illustrates the impact of the energy budget $E_{k}^{\mathrm{max}}$
on the learning performance. It shows that the learning performance
improves with the increase of $E_{k}^{\mathrm{max}}$. This is because
higher energy budget can let the device train with more local data
for better learning performance, and can also let the device finish
the local model uploading with less time consumption to satisfy the
maximum latency constraint. It also shows that the learning performance
saturates with the increase of $E_{k}^{\mathrm{max}}$ when $E_{k}^{\mathrm{max}}$
is very large. This is since when $E_{k}^{\mathrm{max}}$ is very
large, the other constraints such as the transmit power constraint
and the maximum latency constraint restrict the amount of synthetic
data that can be downloaded from the server and become the main bottleneck
of the learning performance. 

Furthermore, Fig. \ref{fig:5} shows that as $E_{k}^{\mathrm{max}}$
increases, the learning performance of the schemes without AIGC saturates
earlier than the schemes with AIGC except TDMA+AIGC. The reasons for
this are explained as follows. A larger $E_{k}^{\mathrm{max}}$ can
let the devices upload the local model to the server more quickly
or train the local model in time such that the constraints of the
problem for the schemes without AIGC can be satisfied more easily.
However, since the training data cannot be increased, the learning
performance of the schemes without AIGC is capped. While for the schemes
with AIGC, besides the benefit mentioned above, more synthetic data
can be downloaded from the server for local training with a higher
$E_{k}^{\mathrm{max}}$ such that the learning performance can be
further improved. It also shows that the learning performance improvement
of the proposed NOMA+AIGC scheme compared to other schemes is obvious
and does not change as $E_{k}^{\mathrm{max}}$ varies.

\begin{figure}[!t]
\centering\includegraphics[width=0.42\columnwidth]{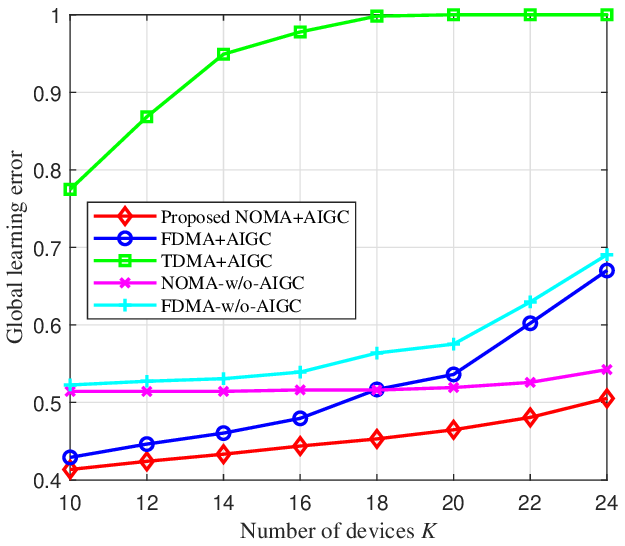}

\caption{Impact of $K$ on the learning performance.\label{fig:6}}
\end{figure}

Fig. \ref{fig:6} illustrates the impact of the number of devices
$K$ on the learning performance. It shows that the learning performance
degrades as $K$ increases. This is due to the fact that under the
maximum latency constraint, more devices lead to a higher probability
of violating this constraint, since the global model broadcasting
rate decreases as $K$ increases and the required time increases.
Besides, more devices also lead to less synthetic data available for
each device, which will degrade the learning performance, since the
objective function in \eqref{eq:p2-o} is a convex function of the
amount of the synthetic data. It also shows that the impact of $K$
on the learning performance of TDMA+AIGC is the severest. This is
because less time resource can be allocated to each device for a larger
$K$ by TDMA+AIGC, and thus the constraints of the problem are much
harder to be satisfied. The proposed NOMA+AIGC scheme is shown to
achieve the best learning performance, and such performance improvement
is still impressive when $K$ is large.

\section{Conclusions\label{sec:Conclusions}}

In this paper, AIGC and NOMA are jointly adopted to enhance the WFL
performance. The synthetic data distribution, two-way communication
and computation resource allocation are jointly optimized to minimize
the global learning error, under various system constraints. Specifically,
an efficient low-complexity local optimal solution to the problem
with partial closed-form results is proposed based on the BCD method
and the analytical method. Extensive simulation results verify the
superiority of the proposed NOMA+AIGC-enhanced scheme compared to
the existing and benchmark schemes such as FDMA or TDMA+AIGC-enhanced
schemes, under various system configurations. Our results have demonstrated
the effectiveness of jointly combining NOMA and AIGC to enhance the
WFL performance.

\appendix{}

\subsection{Proof of Lemma 1\label{subsec:Proof-of-Lemma1}}

Suppose that the optimal computing frequency allocation is $\mathbf{f}^{*}=\{f_{k}^{*},k=1,\ldots,K\},$
where the constraint \eqref{eq:p1-c4} is satisfied with strict inequality
by $f_{j}^{*}$ for a given $j,$ i.e., $f_{j}^{*}>\frac{w\tau\left(D_{j}^{\mathrm{loc}}+D_{j}^{\mathrm{gen}}\right)}{T^{\mathrm{loc}}}.$
Then, we consider another computing frequency allocation $\mathbf{f}^{\star}=\{f_{k}^{\star},k=1,\ldots,K\}$
with $f_{k}^{\star}=f_{k}^{*},k\neq j$ and $f_{j}^{\star}=\frac{w\tau\left(D_{j}^{\mathrm{loc}}+D_{j}^{\mathrm{gen}}\right)}{T^{\mathrm{loc}}}$.
It is clear that $f_{j}^{\star}<f_{j}^{*}$ and $T_{j}^{\mathrm{loc}}=T^{\mathrm{loc}}$
with $f_{j}=f_{j}^{\star}.$ Since $\mathbf{f}$ is only related with
the constraints \eqref{eq:p1-c3}, \eqref{eq:p1-c4}, \eqref{eq:p1-c11},
and \eqref{eq:p1-c17} in the problem \eqref{eq:p1}, it can be easily
verified that $\mathbf{f}^{\star}$ is a feasible solution to the
problem \eqref{eq:p1} and achieves the same objective function value
as the optimal solution $\mathbf{f}^{*}$. This means that the solution
$\mathbf{f}^{\star}$ is also optimal. From \eqref{eq:Ek_loc} and
\eqref{eq:Ek}, it follows that $E_{j}$ with $f_{j}=f_{j}^{\star}$
is smaller than that with $f_{j}=f_{j}^{*}$. This means that $\mathbf{f}^{\star}$
has lower energy consumption than $\mathbf{f}^{*}$. Thus, the optimal
solution $\mathbf{f}^{\star}$ is more desirable than $\mathbf{f}^{*}$.
This completes the proof.

\subsection{Proof of Lemma 2\label{subsec:Proof-of-Lemma2}}

Suppose that the optimal time allocation solution is $\hat{\mathbf{T}}=\{\hat{T}^{\mathrm{down}},\hat{T}^{\mathrm{br}},\hat{T}^{\mathrm{loc}},\hat{T}^{\mathrm{up}}\},$
where $T^{\mathrm{syn}}+\hat{T}^{\mathrm{down}}+N\left(\hat{T}^{\mathrm{br}}+\hat{T}^{\mathrm{loc}}+\hat{T}^{\mathrm{up}}\right)<T^{\mathrm{max}}.$
Then, we construct another time allocation solution $\tilde{\mathbf{T}}=\{\tilde{T}^{\mathrm{down}},\tilde{T}^{\mathrm{br}},\tilde{T}^{\mathrm{loc}},\tilde{T}^{\mathrm{up}}\}$
with $\tilde{\mathbf{T}}=\hat{\mathbf{T}}$ except that $\tilde{T}^{\mathrm{loc}}=\frac{T^{\mathrm{max}}-T^{\mathrm{syn}}-\hat{T}^{\mathrm{down}}}{N}-\hat{T}^{\mathrm{br}}-\hat{T}^{\mathrm{up}}$.
It is shown that the \eqref{eq:p1-c2} is satisfied with equality
by $\tilde{\mathbf{T}}$ and $\tilde{T}^{\mathrm{loc}}>\hat{T}^{\mathrm{loc}}.$
It can be shown that all the remaining constraints of the problem
\eqref{eq:p3} are still satisfied by the solution $\tilde{\mathbf{T}},$
and the objective function value achieved by $\tilde{\mathbf{T}}$
is the same as the optimal solution $\hat{\mathbf{T}}.$ Thus, $\tilde{\mathbf{T}}$
is not only a feasible solution but also an optimal solution. From
\eqref{eq:new-p1-c11}, it is shown that the energy consumption of
each device achieved by the solution $\tilde{\mathbf{T}}$ is lower
than that achieved by the solution $\hat{\mathbf{T}},$ which means
that the optimal solution $\tilde{\mathbf{T}}$ is more desirable
than the optimal solution $\hat{\mathbf{T}}.$ This completes the
proof.

\subsection{Proof of Lemma 3\label{subsec:Proof-of-Lemma3}}

Suppose that the optimal time allocation solution is $\hat{\mathbf{T}}=\{\hat{T}^{\mathrm{down}},\hat{T}^{\mathrm{br}},\hat{T}^{\mathrm{loc}},\hat{T}^{\mathrm{up}}\},$
where $\hat{T}^{\mathrm{down}}R_{j}^{\mathrm{down}}>\Gamma D_{j}^{\mathrm{gen}}$
for $j=\arg\max_{k}\frac{D_{k}^{\mathrm{gen}}}{R_{k}^{\mathrm{down}}}$.
This means that $\hat{T}^{\mathrm{down}}R_{k}^{\mathrm{down}}>\Gamma D_{k}^{\mathrm{gen}},\forall k.$
Then, we construct another solution $\tilde{\mathbf{T}}=\{\tilde{T}^{\mathrm{down}},\tilde{T}^{\mathrm{br}},\tilde{T}^{\mathrm{loc}},\tilde{T}^{\mathrm{up}}\}$
with $\tilde{\mathbf{T}}=\hat{\mathbf{T}}$ except that $\tilde{T}^{\mathrm{down}}=\Gamma\max_{k}\frac{D_{k}^{\mathrm{gen}}}{R_{k}^{\mathrm{down}}},\tilde{T}^{\mathrm{loc}}=\hat{T}^{\mathrm{loc}}+\hat{T}^{\mathrm{down}}-\tilde{T}^{\mathrm{down}}.$
This means that $\tilde{T}^{\mathrm{down}}R_{j}^{\mathrm{down}}=\Gamma D_{j}^{\mathrm{gen}},\tilde{T}^{\mathrm{down}}R_{k}^{\mathrm{down}}>\Gamma D_{k}^{\mathrm{gen}},k\neq j$,
$\tilde{T}^{\mathrm{down}}<\hat{T}^{\mathrm{down}}$ and $\tilde{T}^{\mathrm{loc}}>\hat{T}^{\mathrm{loc}}.$
It can be verified that $\tilde{\mathbf{T}}$ satisfies all the constraints
of the problem \eqref{eq:p3}, and achieves the same objective function
value as the optimal solution $\hat{\mathbf{T}}$. Thus, $\tilde{\mathbf{T}}$
is also an optimal solution. Since $\tilde{T}^{\mathrm{loc}}>\hat{T}^{\mathrm{loc}},$
it can be shown from \eqref{eq:new-p1-c11} that the solution $\tilde{\mathbf{T}}$
achieves lower energy consumption compared to $\hat{\mathbf{T}}.$
This means that the optimal solution $\tilde{\mathbf{T}}$ is more
desirable than the optimal solution $\hat{\mathbf{T}}.$ This completes
the proof.

\subsection{Proof of Lemma 6\label{subsec:Proof-of-Lemma6}}

Suppose that the optimal solution to the problem \eqref{eq:p14} is
$\hat{\mathbf{p}}=\{\hat{p}_{k},k=1,\ldots,K\},\hat{\eta},$ where
$B\log_{2}\left(1+\frac{h_{\hat{k}}\hat{p}_{\hat{k}}}{\sigma^{2}B+h_{\hat{k}}\sum_{j>\hat{k}}\hat{p}_{j}}\right)>\eta D_{\hat{k}}^{\mathrm{gen}}$
for a given $\hat{k}$, and $B\log_{2}\left(1+\frac{h_{k}\hat{p}_{k}}{\sigma^{2}B+h_{k}\sum_{j>k}\hat{p}_{j}}\right)=\eta D_{k}^{\mathrm{gen}}$
for $k\neq\hat{k}$. Then, we can consider another solution $\tilde{\mathbf{p}}=\{\tilde{p}_{k},k=1,\ldots,K\},\tilde{\eta}$
with $\tilde{\mathbf{p}}=\hat{\mathbf{p}},\tilde{\eta}=\hat{\eta}$,
except that $\tilde{p}_{k}$ is determined by $B\log_{2}\left(1+\frac{h_{k}\tilde{p}_{k}}{\sigma^{2}B+h_{k}\sum_{j>k}\tilde{p}_{j}}\right)=\eta D_{k}^{\mathrm{gen}}$.
It is shown that $\tilde{p}_{k}<\hat{p}_{k}$ and thus all the constraints
of the problem \eqref{eq:p14} are satisfied by the solution $\tilde{\mathbf{p}},\tilde{\eta}$.
Therefore, the solution $\tilde{\mathbf{p}},\tilde{\eta}$ is a feasible
solution to the problem \eqref{eq:p14} and consumes less power than
the optimal solution, which indicates that the solution $\tilde{\mathbf{p}},\tilde{\eta}$
is more desirable than the optimal solution $\hat{\mathbf{p}},\hat{\eta}.$
This completes the proof.

\subsection{Proof of Theorem \ref{thm:3}\label{subsec:Proof-ofTheorem1}}

We consider two devices $\hat{k}$ and $\tilde{k}$ with adjacent
decoding orders and $g_{\hat{k}}q_{\hat{k}}^{\mathrm{max}}<g_{\tilde{k}}q_{\tilde{k}}^{\mathrm{max}}$.
There are two possible decoding orders for the devices $\hat{k}$
and $\tilde{k}$, i.e., the device $\hat{k}$ is decoded first or
the device $\tilde{k}$ is decoded first. Let $I$ denote the interference
caused by the devices $\hat{k}$ and $\tilde{k}$ to the devices whose
decoding orders are smaller than the devices $\hat{k}$ and $\tilde{k}$,
and $I^{'}$ denote the interference caused by the devices whose decoding
orders are larger than the devices $\hat{k}$ and $\tilde{k}$ to
the devices $\hat{k}$ and $\tilde{k}$. To guarantee the performance
of other devices, the interference caused by the devices $\hat{k}$
and $\tilde{k}$ is restricted as $g_{\hat{k}}q_{\hat{k}}^{\mathrm{max}}q_{\hat{k}}^{'}+g_{\tilde{k}}q_{\tilde{k}}^{\mathrm{max}}q_{\tilde{k}}^{'}\leq I.$
The problem of optimizing $q_{\hat{k}}^{'}$ and $q_{\tilde{k}}^{'}$
is formulated as
\begin{subequations}
\label{eq:p18}
\begin{alignat}{1}
\max_{q_{\hat{k}}^{'},q_{\tilde{k}}^{'},\theta} & \:\theta\label{eq:p18-o}\\
\mathrm{s.t.} & \:R_{k}^{\mathrm{up}}\geq\theta,k\in\{\hat{k},\tilde{k}\},\label{eq:p18-c1}\\
 & \:g_{\hat{k}}q_{\hat{k}}^{\mathrm{max}}q_{\hat{k}}^{'}+g_{\tilde{k}}q_{\tilde{k}}^{\mathrm{max}}q_{\tilde{k}}^{'}\leq I,\label{eq:p18-c2}\\
 & \:0\leq q_{k}^{'}\leq1,k\in\{\hat{k},\tilde{k}\}.\label{eq:p18-c3}
\end{alignat}
\end{subequations}
Suppose that the device $\hat{k}$ is decoded first. Then, we have
\begin{align}
R_{\hat{k}}^{\mathrm{up}} & =B\log_{2}\left(1+\frac{g_{\hat{k}}q_{\hat{k}}^{\mathrm{max}}q_{\hat{k}}^{'}}{\sigma^{2}B+\sum_{j=1,\pi_{j}>\pi_{\hat{k}}}g_{j}q_{j}^{\mathrm{max}}q_{j}^{'}}\right)\nonumber \\
 & =B\log_{2}\left(\frac{\sigma^{2}B+g_{\hat{k}}q_{\hat{k}}^{\mathrm{max}}q_{\hat{k}}^{'}+g_{\tilde{k}}q_{\tilde{k}}^{\mathrm{max}}q_{\tilde{k}}^{'}+I^{'}}{\sigma^{2}B+g_{\tilde{k}}q_{\tilde{k}}^{\mathrm{max}}q_{\tilde{k}}^{'}+I^{'}}\right),\\
R_{\tilde{k}}^{\mathrm{up}} & =B\log_{2}\left(1+\frac{g_{\tilde{k}}q_{\tilde{k}}^{\mathrm{max}}q_{\tilde{k}}^{'}}{\sigma^{2}B+\sum_{j=1,\pi_{j}>\pi_{\tilde{k}}}g_{j}q_{j}^{\mathrm{max}}q_{j}^{'}}\right)\nonumber \\
 & =B\log_{2}\left(\frac{\sigma^{2}B+g_{\tilde{k}}q_{\tilde{k}}^{\mathrm{max}}q_{\tilde{k}}^{'}+I^{'}}{\sigma^{2}B+I^{'}}\right).
\end{align}
The optimal solution to the problem \eqref{eq:p18} can satisfy the
constraint \eqref{eq:p18-c1} with strict equality, otherwise $q_{\hat{k}}^{'}$
or $q_{\tilde{k}}^{'}$ can be decreased to let the constraint \eqref{eq:p18-c1}
be satisfied with equality to save energy consumption, or $\theta$
can be increased to achieve a higher objective function value. Thus,
from $R_{k}^{\mathrm{up}}=\theta,k\in\{\hat{k},\tilde{k}\},$ we have
\begin{align}
 & q_{\hat{k}}^{'}=\frac{\left(2^{\frac{\theta}{B}}-1\right)\left(\sigma^{2}B+I^{'}\right)2^{\frac{\theta}{B}}}{g_{\hat{k}}q_{\hat{k}}^{\mathrm{max}}},\label{eq:q-hat-1}\\
 & q_{\tilde{k}}^{'}=\frac{\left(2^{\frac{\theta}{B}}-1\right)\left(\sigma^{2}B+I^{'}\right)}{g_{\tilde{k}}q_{\tilde{k}}^{\mathrm{max}}}.\label{eq:q-hat-2}
\end{align}
By inserting \eqref{eq:q-hat-1} and \eqref{eq:q-hat-2} into the
constraints \eqref{eq:p18-c2} and \eqref{eq:p18-c3}, we have
\begin{align}
 & \theta\leq\frac{B}{2}\log_{2}\left(1+\frac{I}{\sigma^{2}B+I^{'}}\right),\label{eq:temp}\\
 & g_{\hat{k}}q_{\hat{k}}^{\mathrm{max}}\geq\left(2^{\frac{\theta}{B}}-1\right)\left(\sigma^{2}B+I^{'}\right)2^{\frac{\theta}{B}},\label{eq:temp-1}\\
 & g_{\tilde{k}}q_{\tilde{k}}^{\mathrm{max}}\geq\left(2^{\frac{\theta}{B}}-1\right)\left(\sigma^{2}B+I^{'}\right).\label{eq:temp-2}
\end{align}
The inequality \eqref{eq:temp} restricts the maximum $\theta$ that
can be supported under the given interference constraint $I,$ while
inequalities \eqref{eq:temp-1} and \eqref{eq:temp-2} also restrict
the maximum $\theta$ that can be supported under the given $g_{\hat{k}}q_{\hat{k}}^{\mathrm{max}}$
and $g_{\tilde{k}}q_{\tilde{k}}^{\mathrm{max}}.$ Since $g_{\hat{k}}q_{\hat{k}}^{\mathrm{max}}<g_{\tilde{k}}q_{\tilde{k}}^{\mathrm{max}},$
the inequality \eqref{eq:temp-1} is tighter than the inequality \eqref{eq:temp-2}.
Thus, the inequality \eqref{eq:temp-2} is dumb, and the maximum $\theta$
can be derived from \eqref{eq:temp} and \eqref{eq:temp-1} accordingly.

Then, assume that the device $\tilde{k}$ is decoded first, and we
get
\begin{align}
R_{\hat{k}}^{\mathrm{up}} & =B\log_{2}\left(1+\frac{g_{\hat{k}}q_{\hat{k}}^{\mathrm{max}}q_{\hat{k}}^{'}}{\sigma^{2}B+\sum_{j=1,\pi_{j}>\pi_{\hat{k}}}g_{j}q_{j}^{\mathrm{max}}q_{j}^{'}}\right)\nonumber \\
 & =B\log_{2}\left(\frac{\sigma^{2}B+g_{\hat{k}}q_{\hat{k}}^{\mathrm{max}}q_{\hat{k}}^{'}+I^{'}}{\sigma^{2}B+I^{'}}\right),\\
R_{\tilde{k}}^{\mathrm{up}} & =B\log_{2}\left(1+\frac{g_{\tilde{k}}q_{\tilde{k}}^{\mathrm{max}}q_{\tilde{k}}^{'}}{\sigma^{2}B+\sum_{j=1,\pi_{j}>\pi_{\tilde{k}}}g_{j}q_{j}^{\mathrm{max}}q_{j}^{'}}\right)\nonumber \\
 & =B\log_{2}\left(\frac{\sigma^{2}B+g_{\hat{k}}q_{\hat{k}}^{\mathrm{max}}q_{\hat{k}}^{'}+g_{\tilde{k}}q_{\tilde{k}}^{\mathrm{max}}q_{\tilde{k}}^{'}+I^{'}}{\sigma^{2}B+g_{\hat{k}}q_{\hat{k}}^{\mathrm{max}}q_{\hat{k}}^{'}+I^{'}}\right).
\end{align}
Similarly, the optimal $q_{\hat{k}}^{'}$ and $q_{\tilde{k}}^{'}$
for the problem \eqref{eq:p18} when the device $\tilde{k}$ is decoded
first shall let the constraint \eqref{eq:p18-c1} be satisfied with
equality. Thus, from $R_{k}^{\mathrm{up}}=\theta,k\in\{\hat{k},\tilde{k}\},$
we get
\begin{align}
 & q_{\hat{k}}^{'}=\frac{\left(2^{\frac{\theta}{B}}-1\right)\left(\sigma^{2}B+I^{'}\right)}{g_{\hat{k}}q_{\hat{k}}^{\mathrm{max}}},\label{eq:q-hat-3}\\
 & q_{\tilde{k}}^{'}=\frac{\left(2^{\frac{\theta}{B}}-1\right)\left(\sigma^{2}B+I^{'}\right)2^{\frac{\theta}{B}}}{g_{\tilde{k}}q_{\tilde{k}}^{\mathrm{max}}}.\label{eq:q-hat-4}
\end{align}
By substituting \eqref{eq:q-hat-3} and \eqref{eq:q-hat-4} into the
constraints \eqref{eq:p18-c2} and \eqref{eq:p18-c3}, we get
\begin{align}
 & \:\theta\leq\frac{B}{2}\log_{2}\left(1+\frac{I}{\sigma^{2}B+I^{'}}\right),\label{eq:temp2}\\
 & \:g_{\hat{k}}q_{\hat{k}}^{\mathrm{max}}\geq\left(2^{\frac{\theta}{B}}-1\right)\left(\sigma^{2}B+I^{'}\right),\label{eq:temp2-1}\\
 & \:g_{\tilde{k}}q_{\tilde{k}}^{\mathrm{max}}\geq\left(2^{\frac{\theta}{B}}-1\right)\left(\sigma^{2}B+I^{'}\right)2^{\frac{\theta}{B}}.\label{eq:temp2-2}
\end{align}
It is noted that the expressions \eqref{eq:temp} and \eqref{eq:temp2}
are the same. It is shown the constraint \eqref{eq:temp2-1} is looser
than the constraint \eqref{eq:temp-1}, while the constraint \eqref{eq:temp2-2}
is also looser than the constraint \eqref{eq:temp-1} due to the fact
that $g_{\hat{k}}q_{\hat{k}}^{\mathrm{max}}<g_{\tilde{k}}q_{\tilde{k}}^{\mathrm{max}}$.
Thus, the maximum allowable $\theta$ under the constraints \eqref{eq:temp2},
\eqref{eq:temp2-1} and \eqref{eq:temp2-2} when the device $\tilde{k}$
is decoded first is equal to or larger than that when the device $\hat{k}$
is decoded first. Therefore, the optimal decoding order is decoding
the device $\tilde{k}$ first.

In the above, we have proved that the optimal decoding order for the
two devices $\hat{k}$ and $\tilde{k}$ with adjacent decoding orders
and $g_{\hat{k}}q_{\hat{k}}^{\mathrm{max}}<g_{\tilde{k}}q_{\tilde{k}}^{\mathrm{max}}$
is decoding the device $\tilde{k}$ first. For multiple devices, since
any two adjacent devices shall satisfy this decoding criterion, we
can conclude that the optimal decoding order is in the descending
order of $g_{k}q_{k}^{\mathrm{max}}.$ This completes the proof.

\subsection{Proof of Lemma 7\label{subsec:Proof-of-Lemma7}}

Suppose that a feasible solution to the problem \eqref{eq:p14} satisfies
$R_{\hat{k}}^{\mathrm{up}}>\theta$ for a given $\hat{k}$. Then,
we can easily decrease the value of $q_{k}^{'}$ until $R_{\hat{k}}^{\mathrm{up}}=\theta$
such that all the constraints are still satisfied. Thus, the constraint
\eqref{eq:p17-c1} can be satisfied with strict equality by the feasible
solution to the problem \eqref{eq:p14}. This completes the proof.

\subsection{Proof of Proposition 1\label{subsec:Proof-of-Proposition1}}

Let $F(\mathbf{D}^{\mathrm{gen}},\mathbf{T},\mathbf{p},\mathbf{q},\boldsymbol{\pi})$
denote the objective function in \eqref{eq:p2-o}. Then, $F(\mathbf{D}^{\mathrm{gen}},\mathbf{T},\mathbf{p},\mathbf{q},\boldsymbol{\pi})$
in the $l$-th iteration of the BCD method is given by
\begin{align}
 & F(\mathbf{D}^{\mathrm{gen}}(l),\mathbf{T}(l),\mathbf{p}(l),\mathbf{q}(l),\boldsymbol{\pi}(l))\nonumber \\
\leq & F(\mathbf{D}^{\mathrm{gen}}(l-1),\mathbf{T}(l-1),\mathbf{p}(l),\mathbf{q}(l),\boldsymbol{\pi}(l))\nonumber \\
\leq & F(\mathbf{D}^{\mathrm{gen}}(l-1),\mathbf{T}(l-1),\mathbf{p}(l-1),\mathbf{q}(l-1),\boldsymbol{\pi}(l-1)).\label{eq:BCD-convergence}
\end{align}
The first inequality in \eqref{eq:BCD-convergence} holds because
we optimally solve the problem of optimizing $\mathbf{D}^{\mathrm{gen}},\mathbf{T}$
with given $\mathbf{p},\mathbf{q},\boldsymbol{\pi}$, and $\mathbf{D}^{\mathrm{gen}}(l),\mathbf{T}(l)$
optimally maximizes the objective function as compared to $\mathbf{D}^{\mathrm{gen}}(l-1),\mathbf{T}(l-1)$.
The second inequality in \eqref{eq:BCD-convergence} holds because
we optimally solve the problem of optimizing $\mathbf{p},\mathbf{q},\boldsymbol{\pi}$
with given $\mathbf{D}^{\mathrm{gen}},\mathbf{T},$ and $\mathbf{p}(l),\mathbf{q}(l),\boldsymbol{\pi}(l)$
optimally maximizes the objective function as compared to $\mathbf{p}(l-1),\mathbf{q}(l-1),\boldsymbol{\pi}(l-1).$
 Since $F(\mathbf{D}^{\mathrm{gen}},\mathbf{T},\mathbf{p},\mathbf{q},\boldsymbol{\pi})$
is clearly lower-bounded, the BCD method used for iteratively solving
the problem \eqref{eq:p2} converges to a local optimal solution.
This completes the proof.

\bibliographystyle{IEEEtran}
\bibliography{bibfile}

\end{document}